\documentclass[reprint,
superscriptaddress,
preprintnumbers,
 amsmath,amssymb,
 aps,
prl,
longbibliography,
]{revtex4-2}
\usepackage{CJKutf8}
\usepackage{multirow}
\usepackage{natbib}
\usepackage{mathtools}
\usepackage{xcolor}
\usepackage{graphicx}
\usepackage{dcolumn}
\usepackage{bbm}
\usepackage{hyperref}

\usepackage{feynmp-auto}
\usepackage{subcaption}

\DeclareMathAlphabet\mathbfcal{OMS}{cmsy}{b}{n}

\newcommand{\eq}{\begin{equation}}
\newcommand{\eqe}{\end{equation}}
\newcommand{\eqa}{\begin{eqnarray}}
\newcommand{\eqae}{\end{eqnarray}}
\newcommand{\nn}{\nonumber}
\newcommand{\bn}{\begin{enumerate}}
\newcommand{\en}{\end{enumerate}}
\def\beq#1\eeq{\begin{align}#1\end{align}}

\def\CO{{\mathcal O}}

\def\CA{{\mathcal A}}

\def\CO{{\mathcal O}}

\newcommand{\bfig}{\begin{figure}}
\newcommand{\efig}{\end{figure}}

\def\bl#1\el{\begin{align} #1 \end{align}}
\def\bg#1\eg{\begin{gather} #1 \end{gather}}

\def\bld#1\eld{\begin{aligned} #1 \end{aligned}}
\def\bgd#1\egd{\begin{gathered} #1 \end{gathered}}

\def\topbotatom#1{\hbox{\hbox to 0pt{$#1\bot$\hss}$#1\top$}} 

\def\bubatom#1{\hbox{\hbox to 1pt{$#1($\hss}$#1)$}}

\newcommand{\RN}[1]{%
  \textup{\uppercase\expandafter{\romannumeral#1}}%
}

\newcommand{\be}{\begin{equation}}
\newcommand{\ee}{\end{equation}}
\newcommand{\ba}{\begin{align}}
\newcommand{\ea}{\end{align}}
\newcommand{\bi}{\begin{itemize}}
\newcommand{\ei}{\end{itemize}}

\let\a=\alpha \let\b=\beta \let\g=\gamma  \let\e=\epsilon
  \let\th=\theta  \let\k=\kappa
\let\l=\lambda \let\m=\mu \let\n=\nu   \let\r=\rho 
\let\s=\sigma     
    \let\D=\Delta

\makeatletter
\newcommand*{\Rom}[1]{\expandafter\@slowromancap\romannumeral #1@}
\makeatother

\makeatletter
\newcommand*{\rom}[1]{\expandafter\romannumeral #1}
\makeatother

\let\nn=\nonumber

\def\beq#1\eeq{\begin{align}#1\end{align}}

\newcommand{\dbar}{
    d\kern-.20em\makebox[0pt][l]{$\bar{}$}\kern.20em
}
\newcommand{\deltabar}{
    \delta\kern-.20em\makebox[0pt][l]{$\bar{}$}\kern.20em
}

\def\fmfblobOval#1#2{\fmfcmd{vblobOval ((#1), \fmfpfx{#2});}}

\begin{document}
\begin{CJK*}{UTF8}{mj}
\title{Radiation eikonal for post-Minkowskian observables
}
\author{Jung-Wook Kim (김정욱)}
\email{jung-wook.kim@aei.mpg.de}
\affiliation{Max Planck Institute for Gravitational Physics (Albert Einstein Institute), 
Am M\"uhlenberg 1, 
D-14476 Potsdam, Germany}

\begin{abstract}
A recent proposal reinterprets the eikonal as the scattering generator, which computes scattering observables through an action as a symmetry generator. The aim of this study is to incorporate dissipative effects from radiation into this framework, where the eikonal is generalised to the \emph{radiation eikonal} by including mediator field degrees of freedom. The proposed generalisation is tested through several post-Minkowskian scattering observables; scattering waveform, radiated momentum, and (time asymmetric) radiation loss in the impulse.
\end{abstract}

\maketitle
\end{CJK*}

\section{Introduction}
The year 2025 marks the tenth anniversary of the first direct detection of gravitational waves; the event GW150914~\cite{LIGOScientific:2016aoc}. Since then, the field has been gradually progressing to a precision science, partly benefiting from improved accuracy of waveform models on the theoretical side. The reinvigoration of interest in the post-Minkowskian (PM) expansion (of two-body dynamics) can partly be attributed to the need for more accurate waveform models~\cite{Damour:2016gwp,Damour:2017zjx}, as binary dynamics is one of the theoretical inputs in building them.

The PM expansion can be viewed as systematic inclusion of general relativistic corrections to special relativistic dynamics, which uses the gravitational constant $G$ as the (dimensionful) expansion parameter~\cite{Westpfahl:1979gu}. In other words, the PM expansion can be viewed as resumming velocity corrections in the post-Newtonian expansion. The natural setup for the PM expansion is scattering dynamics, and due to simplifying features of scattering dynamics such as its gauge-invariant description, the field has recently attracted much attention, especially from the particle physics community~\cite{Bern:2019nnu,Bern:2019crd,Kalin:2020mvi,Kalin:2020fhe,Bern:2021dqo,Dlapa:2021npj,Jakobsen:2022psy,Kalin:2022hph,Dlapa:2022lmu,Dlapa:2024cje,Driesse:2024xad,Bern:2024adl,Driesse:2024feo,Bjerrum-Bohr:2022blt,Kosower:2022yvp,Buonanno:2022pgc}.

One important feature of PM dynamics is the existence of the eikonal. The eikonal is a scalar function that compactly encodes the PM dynamics~\footnote{The existence of the eikonal is proven for conservative dynamics, and this work addresses inclusion of radiative effects. Note that the eikonal may not be single-valued. Well-definedness of its derivatives suffices, since the eikonal itself is not an observable quantity; only though its difference the eikonal represents observable quantities, even for the quantities that cannot be computed from the scattering generator equation [Eq.\eqref{eq:scgen_eq}] such as the time delay~\cite{Shapiro:1964uw,Camanho:2014apa} and polarisation rotation~\cite{Chen:2022clh,Kim:2022iub}.}. This is an advantage from a phenomenological viewpoint, since practical waveform models involve resummation of perturbative two-body dynamics to improve accuracy~\cite{Buonanno:1998gg,Pompili:2023tna,Buonanno:2024byg}; if the dynamics can be compactly encoded by the eikonal, then the study of accurate resummation schemes can be guided by the study of singularities of the eikonal. Some examples in this direction are the studies on spin effect resummation~\cite{Kim:2024grz,Chen:2024bpf}.

The most modern understanding of the (classical) eikonal is viewing it as the scattering generator~\cite{Kim:2024grz,Kim:2024zsf,Kim:2024svw}. In this viewpoint, the space of in-states and out-states of scattering dynamics are considered as isomorphic phase space of asymptotically-free particles, and the eikonal is understood as a symmetry generator that maps in-states to out-states by a canonical transform. The out-state observables $O_{\text{out}}$ are obtained from the in-state observables $O_{\text{in}}$ (corresponding to initial conditions) by the \emph{scattering generator equation}~\cite{Gonzo:2024zxo,Kim:2024grz,Kim:2024zsf,Kim:2024svw},
\begin{align}
\begin{aligned}
    O_{\text{out}} &= e^{\{ \chi , \bullet \}} [O_{\text{in}}]
    \\ &= O_{\text{in}} + \{ \chi , O_{\text{in}} \} + \frac{1}{2!} \{ \chi , \{ \chi , O_{\text{in}} \}\}
    \\ &\phantom{=asdf} + \frac{1}{3!} \{ \chi , \{ \chi , \{ \chi , O_{\text{in}} \}\}\}
    + \cdots \,. \label{eq:scgen_eq}
\end{aligned}
\end{align}
This equation can be viewed as a ``de-quantised'' version of the Kosower-Maybee-O'Connell (KMOC) formalism~\cite{Kosower:2018adc} written in terms of the logarithm of the $S$-matrix~\cite{Lehmann:1957zz,Damgaard:2021ipf,Damgaard:2023ttc}.

This is a novel viewpoint. Conventionally, the eikonal is understood as a saddle-point approximation to the $2 \to 2$ scattering amplitude~\cite{DiVecchia:2023frv}. 
There are two definitions of the impact parameter in the conventional picture; the actual impact parameter $b^\m$ and the ``eikonal impact parameter'' $b_{\text{eik}}^\m$ which is rotated from $b^\m$ by half of the scattering angle, where the ``frame rotation'' compensates for the shift in the saddle-point when computing scattering observables~\cite{DiVecchia:2023frv,Georgoudis:2023eke,Luna:2023uwd}. In the new picture, $b_{\text{eik}}^\m$ is \emph{never} introduced and the same ``rotation effect'' is understood as contributions from nested Poisson brackets~\cite{Kim:2024grz,Kim:2024zsf,Kim:2024svw}; see the discussions around Eq.\eqref{eq:chi1_bracket}.

However, the discussion of the eikonal as the scattering generator has been limited to elastic/conservative $2 \to 2$ scattering. For example, Ref.~\cite{Kim:2024svw} found that the 3PM eikonal computed from the Magnus series correctly reproduces the elastic part (which consists of conservative and radiation reaction contributions) of the impulse, but the momentum lost by radiation is lost. Inspired from eikonal-based approaches to inelastic scattering~\cite{DiVecchia:2022nna,DiVecchia:2022piu,Georgoudis:2023eke,Aoude:2024jxd},
the aim of this study is to formalise a generalisation of the eikonal to include effects from radiation in the scattering generator viewpoint, which can be called the \emph{radiation eikonal}, and show how it can reproduce scattering observables related to effects of radiation.

All computations in this letter will be based on the worldline quantum field theory (WQFT) formalism~\cite{Mogull:2020sak}, and classical limit is implied unless stated otherwise. Mostly negative metric signature is adopted and dimensional regularisation $D = 4 - 2\e$ is implied. The packages \texttt{xTras} of the \texttt{xAct} bundle~\cite{xAct} and \texttt{LiteRed2}~\cite{Lee:2012cn} were used to perform calculations.


\section{Motivating the radiation eikonal}

The radiation eikonal is motivated by the insights of Ref.~\cite{Kim:2024svw}, which is based on three observations: 1) In the WQFT formalism, the impulse is computed as tree diagrams with one marked vertex representing a derivative acting on the vertex, and all propagator causality prescriptions flow towards the marked vertex~\cite{Jakobsen:2022psy}. 2) The Poisson bracket between two vertices (with respect to background worldline variables $X_i^\mu$ and $P_i^\mu$) can be computed as \emph{causality cuts}, where the vertices are joined by retarded-minus-advanced combination of Green's functions for worldline fluctuation fields. The Poisson bracket between two diagrams can be computed as a sum over all possible causality cuts between the diagrams~\cite{Kim:2024grz,Kim:2024svw}. 3) The scattering generator equation [Eq.\eqref{eq:scgen_eq}] defines the eikonal, and the eikonal is reverse-engineered by subtracting nested bracket contributions (diagrammatically computed as causality cuts) from the diagrams for the impulse, and then taking the antiderivative of the remainder. The eikonal computed from the Magnus series matches this definition of the eikonal~\cite{Kim:2024svw}.

One insight of Ref.~\cite{Kim:2024svw} is that this logical flow can be inverted; the eikonal is \emph{defined} by the Magnus series and generates diagrams corresponding to the impulse through the scattering generator equation, when Poisson brackets are diagrammatically computed as causality cuts. 
Another insight is that causality cuts generalise to any two-point functions, since they are a special case of the Peierls bracket~\cite{Peierls:1952cb} which applies to general field theories. 
The Magnus series treats worldline propagators and graviton propagators on an equal footing, 
therefore this definition of the eikonal solves the problem of determining causality prescriptions for graviton propagators, 
which led to an IR-finite result for the 3PM eikonal~\cite{Kim:2024svw}.

However, the 3PM eikonal computed from the Magnus series failed to account for radiation loss in the impulse, which corresponds to dissipative effects from bremsstrahlung~\cite{Kim:2024svw}. 
The deficit was conjectured to arise from neglecting graviton degrees of freedom in the Poisson brackets. 
In other words, the vertices in the original WQFT setup could not be joined by graviton causality cuts, 
and resulted in missing diagrams when the impulse was computed by the scattering generator equation.

This problem has a simple solution; introduce new vertices that can be joined by graviton causality cuts, and add diagrams with these vertices to the eikonal. 
The augmented eikonal will generate extra diagrams corresponding to Poisson bracket contributions from background graviton fields, and complete the set of diagrams for the scattering observable of interest. 
The radiation eikonal formalises this idea and expands the set of observables that can be computed within the eikonal framework, 
by generating radiative observables through the scattering generator equation.


\section{Radiation eikonal and symplectic structure of graviton phase space}

The basic setup of WQFT is to consider straight-line particle trajectories on a flat background metric. The formalism proceeds by considering quantised fluctuations on these background variables; the graviton field $h_{\m\n} (x)$ and the worldline fluctuations $z_i^\m (\s_i)$. The metric $g_{\m\n} (x)$ and the worldline $x_i^\m (\s_i)$ are expanded as ($\k = \sqrt{32\pi G}$)
\begin{align}
\begin{aligned}
    g_{\m\n} (x) &= g_{\m\n}^{(0)} (x) + \k h_{\m\n} (x) \,,\quad g_{\m\n}^{(0)} (x) = \eta_{\m\n} \,,
    \\ x_i^\m (\s_i) &= X_i^\m + v_i^\m \s_i + z_i^\m (\s_i) \,,\quad v_i^\m := \frac{P_i^\m}{m_i} \,.
\end{aligned}
\end{align}
The worldline $x_i^\m (\s_i)$ describes a particle of mass $m_i$ with momentum $P_i^\m$ (or velocity $v_i^\m$) passing through $X_i^\m$ at worldline time $\s_i = 0$, satisfying the on-shell condition $P_i^2 = m_i^2$. 
The action is given by gauge-fixed Einstein-Hilbert term and gauge-fixed Polyakov-type point-particle action,
\begin{align}
\begin{aligned}
    S &= 
    S_{\text{EH}} + S_{\text{gf}}
    + \sum_{i = 1,2} - \frac{m_i}{2} \int g_{\m\n} \frac{d x_i^\m}{d \s_i} \frac{ d x_i^\n}{d \s_i} d\s_i \,,
\end{aligned}
\end{align}
where the graviton field is in linearised harmonic gauge; see Ref.~\cite{Mogull:2020sak} for details and Feynman rules. 

The eikonal $\frac{i}{\hbar} \chi$ is defined as the vacuum expectation value of the Magnus series of the interaction Hamiltonian $\frac{1}{i \hbar} H_{\text{int}}$, where all fluctuation fields ($z_i^\m$ and $h_{\m\n}$) are integrated out~\cite{Kim:2024svw}. The resulting eikonal is a function of the background variables $X_i^\m$ and $P_i^\m$. 

The \emph{radiation eikonal} is proposed as the vacuum expectation value of the Magnus series when the background metric is changed from flat to weak graviton field,
\begin{align}
    g^{(0)}_{\m\n} (x) = \eta_{\m\n} && \to && g^{(0)}_{\m\n} (x) = \eta_{\m\n} + \k H_{\m\n} (x) \,.
\end{align}
The background graviton field $H_{\m\n}$ is assumed to satisfy the same gauge conditions as the graviton fluctuation field $h_{\m\n}$, although it is not assumed to satisfy on-shell conditions. Only $h_{\m\n}$ is integrated out when computing the radiation eikonal, which depends on the background variables $X_i^\m$, $P_i^\m$, and $H_{\m\n}$.

While generalisation of the eikonal to nontrivial backgrounds was considered in the literature~\cite{Adamo:2021rfq}, the approach advocated here differs in that the background graviton field is introduced for the \emph{dynamics} of gravitons; the main purpose of the generalisation is to include Poisson brackets of graviton fields into the scattering generator equation [Eq.\eqref{eq:scgen_eq}], where the background graviton field $H_{\m\n}$ is set to zero at the last stage of computations to reflect radiation-less initial conditions.

The defining Poisson brackets are
\begin{align}
\begin{aligned}
    \{ X_i^\m , P_j^\n \} &= - \delta_{ij} \eta^{\m\n} \,,
    \\ \{ H_{\m\n} (x) , H_{\a\b} (y) \} &= - G^{(R-A)}_{\m\n;\a\b} (x; y) \,, 
\end{aligned} \label{eq:defPB}
\end{align}
where $G^{(R-A)}_{\m\n;\a\b} (x; y)$ is the retarded-minus-advanced combination of Green's functions for the graviton fluctuation field $h_{\m\n} (x)$ implementing the graviton causality cut.

In practice, the radiation eikonal is equivalent to computing connected diagrams with external graviton legs, where all external graviton legs are amputated and attached to the background graviton field $H_{\m\n} (x)$.

The dependence on the background graviton field can be nonlinear. For example, the quadratic-in-graviton-field radiation eikonal is computed by summing over connected diagrams with two external graviton legs.

It is convenient to work in momentum space and use $H_{\m\n} (k)$ instead of $H_{\m\n} (x)$~\footnote{The positive frequency expansion $f(x) = \int_k e^{-ikx} f(k)$ and outgoing convention (reading coefficients of $f(-k)$ as vertex rules) are employed to ensure consistency with the Feynman rules given in Ref.~\cite{Mogull:2020sak}.}, where the graviton field Poisson brackets take the form ($\deltabar(x) = 2 \pi \delta (x)$)
\begin{align}
    \{ H_{\m\n} (q) , H_{\a\b} (k) \} &= i P_{\m\n;\a\b} \, \text{sgn} (k^0) \deltabar(k^2) \deltabar^D (q + k) \,.
\end{align}
The projector $P_{\m\n;\a\b}$ follows from the gauge condition (linearised harmonic gauge)~\cite{Mogull:2020sak}.
\begin{align}
    P_{\m\n;\a\b} &= \frac{1}{2} \left[ \eta_{\m\a} \eta_{\n\b} + \eta_{\m\b} \eta_{\n\a} - \frac{2}{D-2} \eta_{\m\n} \eta_{\a\b} \right] \,.
\end{align}


\section{Leading order radiation eikonal}

\begin{figure}[htbp]
    \centering
\begin{subfigure}[t]{0.2\textwidth}
    \centering
    \begin{fmffile}{radeik1}
    \begin{fmfgraph*}(50,40)
        \fmfstraight
        \fmftop{i1,a1,b1,o1}
        \fmfbottom{i2,a2,b2,o2}
        \fmf{dots}{i1,a1}
        \fmf{dots}{b1,o1}
        \fmf{dots}{i2,a2,b2,o2}
        \fmf{plain,width=2}{a1,b1}
        \fmf{phantom}{o1,c1,o2}
         \fmffreeze
        \fmf{wiggly,width=1.5,tension=0}{a1,a2} 
        \fmf{wiggly,width=1.5,tension=0}{b1,c1}
        \fmfv{decor.shape=circle,decor.size=4}{a1,a2,b1}
    \end{fmfgraph*}
    \end{fmffile}
    \caption{}
\label{fig:LOradeika}
\end{subfigure}
\begin{subfigure}[t]{0.2\textwidth}
    \centering
    \begin{fmffile}{radeik2}
    \begin{fmfgraph*}(50,40)
        \fmfstraight
        \fmfbottom{i1,a1,b1,o1}
        \fmftop{i2,a2,b2,o2}
        \fmf{dots}{i1,a1}
        \fmf{dots}{b1,o1}
        \fmf{dots}{i2,a2,b2,o2}
        \fmf{plain,width=2}{a1,b1}
        \fmf{phantom}{o1,c1,o2}
         \fmffreeze
        \fmf{wiggly,width=1.5,tension=0}{a1,a2} 
        \fmf{wiggly,width=1.5,tension=0}{b1,c1} 
        \fmfv{decor.shape=circle,decor.size=4}{a1,a2,b1}
    \end{fmfgraph*}
    \end{fmffile}
    \caption{}
\label{fig:LOradeikb}
\end{subfigure}
\begin{subfigure}[t]{0.2\textwidth}
    \centering
    \begin{fmffile}{radeik3}
    \begin{fmfgraph*}(30,40)
        \fmfstraight
        \fmftop{i1,a1,o1}
        \fmfbottom{i2,a2,o2}
        \fmf{dots}{i1,a1,o1}
        \fmf{dots}{i2,a2,o2}
        \fmf{phantom}{a1,c1,a2}
        \fmf{phantom}{o1,c2,o2}
         \fmffreeze
        \fmf{wiggly,width=1.5,tension=0}{a1,c1,a2} 
        \fmf{wiggly,width=1.5,tension=0}{c1,c2} 
        \fmfv{decor.shape=circle,decor.size=4}{a1,c1,a2}
    \end{fmfgraph*}
    \end{fmffile}
    \caption{}
\label{fig:LOradeikc}
\end{subfigure}
\caption{Diagrams for leading order radiation eikonal. Dotted lines denote background worldlines, solid lines denote propagating worldline fluctuations, and wavy lines denote gravitons. The external graviton legs are amputated and attached to the background field $H_{\m\n}$.}
\label{fig:LOradeik}
\end{figure}
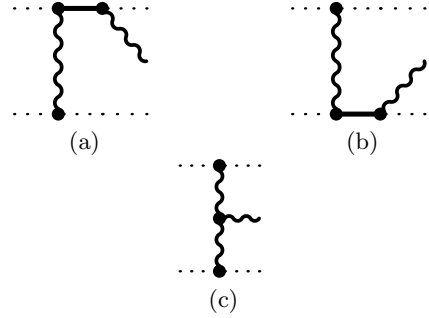

The leading-order (LO) radiation eikonal $\chi_{(1.5)}^{H^1}$ is presented as a concrete example, which is computed as the sum of three diagrams given in Fig.~\ref{fig:LOradeik}~\footnote{A careful reader will argue that the LO contribution is the ``disconnected'' terms of order $\CO(G^{1/2})$ corresponding to the Coulomb background generated by free particles. This contribution has been neglected since it does not seem to affect the dynamics, at least at the LO. See the footnote before Eq.~\eqref{eq:LOradJ} for possible violations of this assumption.},

\begin{widetext}
\begin{align}
\begin{aligned}
    i \chi_{(1.5)}^{H^1} [H (k)] &= i \frac{\k^3 m_1 m_2}{8} \int_{q_1,q_2,k} \hskip -17pt H_{\m\n} (-k) \frac{(2 (k \cdot v_1) v_1^\m \delta^\n_\a - v_1^\m v_1^\n k_\a)(2 (k \cdot v_1) v_1^\r \eta^{\s\a} - v_1^\r v_1^\s q_2^\a)}{(k \cdot v_1)^2} \frac{P_{\r\s;\l\eta} v_2^\l v_2^\eta}{q_2^2} + (1 \leftrightarrow 2)
    \\ &\phantom{=asdf} + i \frac{\k^3 m_1 m_2}{8} \int_{q_1, q_2, k}  \hskip -17pt H_{\m\n} (-k) V_3^{\m\n;\a\b;\r\s} \frac{P_{\a\b;\g\delta} v_1^\g v_1^\delta}{q_1^2} \frac{P_{\r\s;\l\eta} v_2^\l v_2^\eta}{q_2^2}
    \\ &=: i \int \frac{d^D k}{(2\pi)^D} \, \CA_{(1.5)}^{\m\n} (k) H_{\m\n} (-k) \,,
\end{aligned} \label{eq:LOradeik}
\end{align}
where $\frac{i \k}{2} V_3^{\m\n;\a\b;\l\delta}$ is the three-graviton vertex rule and
\begin{align}
    \int_{q_1,q_2,k} &:= \int \frac{d^D q_1}{(2\pi)^D}\frac{d^D q_2}{(2\pi)^D} \frac{d^D k}{(2\pi)^D} \, e^{i (q_1 \cdot X_1 + q_2 \cdot X_2)} \, \deltabar (v_1 \cdot q_1) \deltabar (v_2 \cdot q_2) \deltabar^D (q_1 + q_2 - k) \,,
\end{align}
is the integral measure.
\end{widetext}

The superscript $H^1$ denotes linear dependence on the graviton field and the subscript $(1.5)$ denotes $\CO(G^{3/2})$ power counting. Note that the LO radiation eikonal has almost the same structure as the LO scattering waveform,
\begin{align}
    k^2 \langle h_{\m\n} (k) \rangle_{\text{WQFT}} \stackrel{\cdot}{=} - i \CA_{(1.5) \m\n} (k) \,,
\end{align}
where the LHS is the (sign-flipped~\footnote{The d'Alembertian operator $\square \Leftrightarrow - k^2$ has an extra minus sign in momentum space.}) LO waveform given in Refs.~\cite{Mogull:2020sak,Jakobsen:2021smu} and $\stackrel{\cdot}{=}$ is the equivalence up to propagator $i0^+$ prescriptions~\footnote{There is a distinction made between the $i0^+$ prescription and the causality prescription. The $i0^+$ prescription can be retarded, advanced, or Feynman; the causality prescription can only be retarded or advanced.}.

The omitted causality prescriptions of the propagators in Eq.\eqref{eq:LOradeik} need to be prescribed based on the Magnus series, which is generally different from the symmetric prescription~\cite{Kim:2024svw}. 
The relevant propagators are the worldline fluctuation propagators in Figs.\ref{fig:LOradeika}-\ref{fig:LOradeikb} for the computations presented in this letter, because the internal graviton propagators cannot go on-shell for the considered diagrams. When graviton propagators are ignored in these diagrams, the Magnus series prescribes equal weights to retarded and advanced prescriptions for worldline propagators.


\section{Scattering waveform}
The scattering waveform is computed from the scattering generator equation [Eq.\eqref{eq:scgen_eq}] as
\begin{align}
\begin{aligned}
    H_{\m\n}^{\text{out}} (k) &= e^{\{ \chi , \bullet \}} [H_{\m\n}^{\text{in}} (k)]
    \\ &= H_{\m\n}^{\text{in}} (k) + \{ \chi , H_{\m\n}^{\text{in}} (k) \}
    \\ &\phantom{=a} + \frac{1}{2} \{ \chi , \{ \chi , H_{\m\n}^{\text{in}} (k) \} \} + \cdots \,.
\end{aligned}
\end{align}
The LO scattering waveform $S_{\m\n}^{\text{LO}}$ is given as
\begin{align}
\begin{aligned}
    S_{\m\n}^{\text{LO}} (k) &= \left. \{ \chi_{(1.5)}^{H^1} , H_{\m\n}^{\text{in}} (k) \} \right|_{H_{\m\n}^{\text{in}} = 0}
    \\ &= i P_{\m\n;\a\b} \, \text{sgn} (k^0) \deltabar(k^2) \CA_{(1.5)}^{\a\b} (k) \,, 
\end{aligned} \label{eq:LO_wf}
\end{align}
where the initial condition $H_{\m\n}^{\text{in}} = 0$ is taken \emph{after} evaluating the Poisson brackets. By construction, Eq.\eqref{eq:LO_wf} computes the LO waveform of Refs.~\cite{Mogull:2020sak,Jakobsen:2021smu}, where the $\{ H, H \}$ Poisson bracket imposes the on-shell condition on the radiated graviton.

The next-to-leading-order (NLO) waveform $S_{\m\n}^{\text{NLO}}$ is given as a sum of two terms (since $\{ \chi_{(1)} , H_{\m\n} \} = 0$),
\begin{align}
\begin{aligned}
    S_{\m\n}^{\text{NLO}} (k) &= \{ \chi_{(2.5)}^{H^1} , H_{\m\n}^{\text{in}} (k) \} 
    \\ &\phantom{=a} + \frac{1}{2} \{ \chi_{(1)} , \{ \chi_{(1.5)}^{H^1} , H_{\m\n}^{\text{in}} (k) \} \} \,,
\end{aligned} \label{eq:NLO_wf}
\end{align}
where $\chi_{(1)}$ is the 1PM eikonal and it is understood that the initial condition $H_{\m\n}^{\text{in}} = 0$ is imposed at the end of evaluation. 

The second line of Eq.\eqref{eq:NLO_wf} can be called the ``causality cut contribution'' to the waveform, since causality cuts compute Poisson brackets between subdiagrams~\cite{Kim:2024grz,Kim:2024zsf,Kim:2024svw}.
Similar to the contribution colloquially called the ``KMOC cut''~\cite{Caron-Huot:2023vxl,Georgoudis:2023eke,Georgoudis:2024pdz,Bini:2024rsy} in the KMOC formalism~\cite{Kosower:2018adc,Cristofoli:2021vyo}, this term computes rotation of the LO waveform due to LO impulse, which was one of the effects that lead to initial disagreement between post-Newtonian and PM calculations~\cite{Brandhuber:2023hhy,Herderschee:2023fxh,Elkhidir:2023dco,Georgoudis:2023lgf,Bini:2023fiz}. 

The Poisson bracket $\{ \chi_{(1)} , \bullet \}$ can be recast as a differential operator which rotates all frame-defining vector variables ($b^\m$ and $P_i^\m$) by the LO scattering angle~\footnote{For spinning systems, $\sum_i \{ \chi_{(1)} , S_i^{\m\n} \} \frac{\partial}{\partial S_i^{\m\n}}$ implements precession by the LO spin kick, consistent with the conventional picture of saddle-point shifts~\cite{Luna:2023uwd}. In other words, the $\mathcal{D}_{SL}$ operator introduced in Ref.~\cite{Bern:2020buy} can be viewed as compensation terms for neglecting nested brackets in the scattering generator equation [Eq.\eqref{eq:scgen_eq}] and ``noncommutativity'' of the impact parameter space [Eq.\eqref{eq:bbPB}]; see e.g. the scattering observable computations of Ref.~\cite{Akpinar:2025bkt} where instead of defining $b^\m$ to manifestly satisfy the orthogonality conditions $b \cdot P_i = 0$ [Eq.\eqref{eq:impbdef}], the Dirac brackets constructed from imposing orthogonality conditions as constraints were used. When brackets between variables that manifestly satisfy the constraint conditions are considered, Poisson brackets can be used in place of Dirac brackets~\cite{Kim:2024grz}.},
\begin{align}
    \{ \chi_{(1)} , \bullet \} &= \{ \chi_{(1)} , b^\m \} \frac{\partial}{\partial b^\m} + \sum_{i=1,2} \{ \chi_{(1)} , P_i^\m \} \frac{\partial}{\partial P_i^\m} \,, \label{eq:chi1_bracket}
\end{align}
where
\begin{align}
    \{ \chi_{(1)} , b^\m \} &= - \frac{2 (2\g^2 - 1) G m_1 m_2}{\sqrt{\g^2 - 1}} \left[ \frac{w_1^\m}{m_1} - \frac{w_2^\m}{m_2} \right] \,,
    \\ \{ \chi_{(1)} , P_1^\m \} &= - \frac{2 (2\g^2 - 1) G m_1 m_2}{\sqrt{\g^2 - 1}} \frac{b^\m}{|b^2|} \nn
    \\ &= - \{ \chi_{(1)} , P_2^\m \} \,.
\end{align}
It is important to note that the impact parameter space is ``noncommutative''~\cite{Kim:2024grz,Kim:2024zsf,Kim:2024svw},
\begin{align}
    \{ b^\m , b^\n \} &\stackrel{\cdot}{=} \frac{b^\m w_1^\n - b^\n w_1^\m}{m_1} - \frac{b^\m w_2^\n - b^\n w_2^\m}{m_2} \,, \label{eq:bbPB}
\end{align}
where $\stackrel{\cdot}{=}$ denotes equivalence up to on-shell conditions $P_i^2 = m_i^2$. This bracket follows from the definition of $b^\m$,
\begin{align}
\begin{aligned}
    b^\m &:= X_{12}^\m - w_1^\m \, (v_1 \cdot X_{12}) - w_2^\m \, (v_2 \cdot X_{12}) \,,
    \\ X_{12}^\m &:= X_1^\m - X_2^\m \,,\quad w_{1/2}^\m := \frac{\g \, v_{2/1}^\m - v_{1/2}^\m}{\g^2 -1 } \,.
\end{aligned} \label{eq:impbdef}
\end{align}
Here, $\g := v_1 \cdot v_2$ is the Lorentz factor and $w_i^\m$ are dual velocity vectors satisfying $v_i \cdot w_j = \delta_{ij}$.

Therefore, the ``causality cut contribution'' of the NLO waveform can be written as
\begin{align}
    \frac{1}{2} \left[ \{ \chi_{(1)} , b^\a \} \frac{\partial}{\partial b^\a} + \sum_{i=1,2} \{ \chi_{(1)} , P_i^\a \} \frac{\partial}{\partial P_i^\a} \right] S_{\m\n}^{\text{LO}} (k) \,, \label{eq:NLO_rot_wf}
\end{align}
which corrects the LO waveform $S_{\m\n}^{\text{LO}}$ by a rotation of half of the LO scattering angle.


\section{Radiated momentum}
The gravitational wave momentum is defined as~\cite{Herrmann:2021lqe}
\begin{align}
    P_H^\m &= \int d\Phi_k \, k^\m P_{\a\b;\r\s} H^{\a\b} (k) H^{\r\s} (k) \,, 
\end{align}
where indices are raised with $\eta^{\m\n}$ and 
\begin{align}
    d\Phi_k = \frac{d^D k}{(2\pi)^D} \th(k^0) \deltabar(k^2) \,,
\end{align}
is the Lorentz-invariant phase space measure for the graviton. Imposing the initial condition $H_{\m\n}^{\text{in}} = 0$, the first non-vanishing out-state gravitational wave momentum from the scattering generator equation [Eq.\eqref{eq:scgen_eq}] is
\begin{align}
\begin{aligned}
    P_{H, \text{out}}^\m &= \frac{1}{2} \{ \chi_{(1.5)}^{H^1} , \{ \chi_{(1.5)}^{H^1} , P_H^\m \} \}
    \\ &= \int d\Phi_k \, k^\m P_{\a\b;\r\s} \left[ i \CA_{(1.5)}^{\a\b} (k) \right] \left[ i \CA_{(1.5)}^{\r\s} (k) \right] \,,
\end{aligned} \label{eq:LOradmom}
\end{align}
where ($\stackrel{\cdot}{=}$ denotes equivalence at $D = 4$)
\begin{align}
    P_{\m\n;\a\b} P^{\a\b}_{\phantom{\a\b};\g\delta} P^{\g\delta}_{\phantom{\g\delta};\r\s} \stackrel{\cdot}{=} P_{\m\n;\r\s} \,,
\end{align}
was used to simplify the expression. The expression for $P_{H, \text{out}}^\m$ [Eq.\eqref{eq:LOradmom}] becomes the same integral evaluated in Ref.~\cite{Jakobsen:2021smu} for radiated momentum, which can be expressed using the radiation loss [Eq.\eqref{eq:3PMrl2}] as~\footnote{This relation follows from translation invariance of LO radiation eikonal $\chi_{(1.5)}^{H^1} [H(x)]$ in position space, where $x^\m$ is dual to $k^\m$ in Eq.\eqref{eq:LOradeik}; $\frac{\partial \chi_{(1.5)}^{H^1}}{\partial X_1^\m} + \frac{\partial \chi_{(1.5)}^{H^1}}{\partial X_2^\m} + \frac{\partial \chi_{(1.5)}^{H^1}}{\partial x^\m} = 0$. When reformulated in position space, the integrand of Eq.\eqref{eq:LOradmom} is equivalent to the integrand of Eq.\eqref{eq:3PMrl2} with $\frac{\partial}{\partial X_{1\m}}$ substituted by $\frac{\partial}{\partial x_\m}$, and translation invariance implies $P_{H, \text{out}}^\m + \D_{(3,rl)} P_1^\m + \D_{(3,rl)} P_2^\m = 0$.}
\begin{align}
     P_{H, \text{out}}^\m = - \left[ \D_{(3,rl)} P_1^\m + (1 \leftrightarrow 2) \right] \,.
\end{align}


\section{Radiation loss of impulse}
It was conjectured in Ref.~\cite{Kim:2024svw} that part of the 3PM impulse associated to the momentum lost by radiation, the time-asymmetric \emph{radiation loss} $\D_{(3,rl)} P_1^\m$, can be computed from the radiation eikonal as the following term in the expansion of the scattering generator equation,
\begin{align}
    \D_{(3,rl)} P_1^\m &= \frac{1}{2!} \{ \chi_{(1.5)}^{H^1} , \{ \chi_{(1.5)}^{H^1} , P_1^\m \}\} \,. \label{eq:3PMrl1}
\end{align}
This is indeed the case; the radiation loss in the impulse can be computed from the scattering generator equation [Eq.\eqref{eq:scgen_eq}] when the Poisson brackets of the background graviton fields [Eq.\eqref{eq:defPB}] are taken into account.

The radiation loss [Eq.\eqref{eq:3PMrl1}] can be reorganised as
\begin{widetext}
\begin{align}
\begin{aligned}
    \D_{(3,rl)} P_1^\m &= \frac{1}{2} \int \frac{d^D k}{(2 \pi)^D} \frac{d^D q}{(2 \pi)^D} \left( \frac{\partial \chi_{(1.5)}^{H^1}}{\partial H_{\a\b} (k)} \right) \times \{ H_{\a\b} (k) , H_{\l\s} (q) \} \times \left( \frac{\partial}{\partial H_{\l\s} (q)} \left[ - \frac{\partial \chi_{(1.5)}^{H^1}}{\partial X_{1\m}} \right] \right)
    \\ &= \frac{1}{2} 
    \left[ \quad
    \begin{fmffile}{impulse}
        \fmfcmd{%
            style_def wiggly_arrow expr p = 
                cdraw (wiggly p); 
                shrink (1.5); 
                    cfill (arrow p); 
                endshrink; 
            enddef;}
        \fmfcmd{input vbloboval;}
        \parbox{120pt}{
        \begin{fmfgraph*}(120,40)
            \fmfstraight
            \fmftop{i1,l1,cl1,c1,cr1,r1,o1}
            \fmfbottom{i2,l2,cl2,c2,cr2,r2,o2}
            \fmf{dots}{i1,r1}
            \fmf{plain,width=2}{r1,o1}
            \fmf{dots}{i2,o2}
            \fmf{phantom}{l1,gl1,l2}
            \fmf{phantom}{r1,gr1,r2}
            \fmf{phantom}{c1,gc1,c2}
            \fmffreeze
            \fmf{wiggly_arrow,width=1.5,tension=-0.2}{gl1,gr1}
            \fmfv{decor.shape=cross,decor.size=18}{gc1}
            \fmfblobOval{1.1h}{gl1}
            \fmfblobOval{1.1h}{gr1}
        \end{fmfgraph*}
        }
    \end{fmffile}
    \quad \right]
    \\ &
    = \frac{G^3 m_1^2 m_2^2}{|b^2|^{3/2}} \left[ \frac{\pi}{16} \frac{\g (33 - 112 \g^2 + 165 \g^4 - 70 \g^6)}{(\g^2 - 1)^2} \, \text{arccosh} (\g) \right.
    \\ &\phantom{=asdfasdfas} + \frac{\pi}{8} \frac{(-5 + 76 \g - 150 \g^2 + 60 \g^3 + 35 \g^4)}{(\g^2 - 1)^{1/2}} \log \left( \frac{1 + \g}{2} \right)
    \\ &\phantom{=asdfasdfas} \left. + \frac{\pi}{48} \frac{(-1151 + 3336 \g - 3148 \g^2 + 912 \g^3 - 339 \g^4 +  552 \g^5 - 210 \g^6)}{(\g^2 - 1)^{3/2}} \right] w_2^\m \,,
\end{aligned} \label{eq:3PMrl2}
\end{align}
\end{widetext}
where the left blob represents $\frac{\partial \chi_{(1.5)}}{\partial H}$, the right blob represents $\frac{\partial^2 \chi_{(1.5)}}{\partial H \partial X}$, and the causality cut graviton (wavy line with crossed arrow) represents the $\{ H, H \}$ bracket. This is the only non-vanishing contribution in Eq.\eqref{eq:3PMrl1}, since other terms vanish due to the initial conditions for the background graviton field $H_{\m\n}^{\text{in}} = 0$.

To evaluate Eq.\eqref{eq:3PMrl2}, the momentum space integral was evaluated first and then Fourier transformed to impact parameter space, where the vector integral was projected onto the basis of $q^\m$ (dual to $b^\m$) and $w_{j}^\m$ in momentum space. The momentum space integral was evaluated using reverse unitarity~\cite{Herrmann:2021lqe}, where the integral is evaluated as the difference between two (multi)loop integrals with differing $i0^+$ prescriptions; the causality cut graviton in the middle of the diagram is the difference between retarded and advanced Green's functions. This means only the master integrals that depend on the cut graviton's causality prescription contributes to the integral. 

In the master integral notations of Ref.~\cite{Jakobsen:2023oow}, the relevant master integrals are the $v$-type master integrals $I_{(9-12)}$, which flip their signs when the relevant graviton propagator (corresponding to the causality cut graviton) switches its causality prescription from retarded to advanced. The master integral $I_{(9)}^{\pm}$ depends on the causality prescription of the worldline propagator, which is denoted by its superscript $\pm$. 

The relevant master integrals $I_{(9-12)}$ only appear in the coefficient of $w_2^\m$ when evaluating Eq.\eqref{eq:3PMrl2}; the coefficient of $q^\m$ only depends on master integrals irrelevant for the causality cut, while the coefficient of $w_1^\m$ vanishes identically. 
Due to the symmetric causality prescription imposed by the Magnus series, the master integral $I_{(9)}$ should be inserted as $I_{(9)} = \frac{1}{2} I_{(9)}^+ + \frac{1}{2} I_{(9)}^-$. 
This combination ensures cancellation of IR divergences when computing Eq.\eqref{eq:3PMrl2}, similar to how the causality prescription weighting based on the Magnus series was important for getting an IR-finite result for the 3PM eikonal~\cite{Kim:2024svw}. 

The computed radiation loss to the 3PM impulse in Eq.\eqref{eq:3PMrl2} is the term that was exactly missing in the 3PM impulse computed from the elastic/conservative eikonal up to 3PM in Ref.~\cite{Kim:2024svw}, when compared to the full 3PM impulse given in Ref.~\cite{Jakobsen:2022psy}; the elastic eikonal only computed the conservative and radiation reaction contributions of the impulse.


\section{Discussion}

The radiation eikonal was introduced as a generalisation of the eikonal for describing (dissipative) effects from radiation, and was shown to reproduce scattering observables, at least up to LO in the PM expansion. 
The scattering generator equation [Eq.\eqref{eq:scgen_eq}] correctly reproduced the scattering waveform, radiated momentum, and radiation loss to the impulse from the LO radiation eikonal, and also provided a more coherent understanding of cut contributions at NLO corresponding to ``frame rotation'' of the LO waveform. 
A similar phenomenon observed in the NLO radiation loss~\cite{Heissenberg:2025ocy}, where the cut contribution computes rotation by the LO scattering angle, is also expected to be understood as contributions from nested brackets in the scattering generator viewpoint.

Although the computations were confined to linear momentum in this letter, extension to angular momentum would be straightforward. 
One interesting computation would be the LO radiation loss of angular momentum from the radiation eikonal~\footnote{Eq.~\eqref{eq:LOradJ} assumes vanishing contributions from the ``disconnected'' radiation eikonal $\chi_{(0.5)}^{H^1}$. The LO will become $\CO(G^2)$ instead if this assumption is violated.},
\begin{align}
    \D_{(3,rl)} J_1^{\m\n} = \frac{1}{2!} \{ \chi_{(1.5)}^{H^1} , \{ \chi_{(1.5)}^{H^1} , X_1^\m P_1^\n - X_1^\n P_1^\m \}\} \,, \label{eq:LOradJ}
\end{align}
which may shed a new light on the Bondi–Metzner–Sachs (BMS) frame ambiguities discussed in the literature~\cite{Manohar:2022dea,DiVecchia:2022owy,Mao:2024ryq,Biswas:2024ept}.

Higher PM order dynamics will be rich of more complex physical effects. 
For example, the scattering generator equation implies that the following term contributes to the 4PM impulse,
\begin{align}
    \D_{(4)}P_1^\m \supset \frac{1}{3!} \{ \chi_{(1.5)}^{H^1} , \{ \chi_{(1)} , \{ \chi_{(1.5)}^{H^1} , P_1^\m \} \} \} \,.
\end{align}
This term has an interpretation reminiscent of tail effects; a graviton is radiated ($\{ \chi_{(1.5)}^{H^1} , \bullet \}$), scatters from the binary potential ($\{ \chi_{(1)} , \bullet \}$), and then gets re-absorbed by the binary ($\{ \chi_{(1.5)}^{H^1} , \bullet \}$). 
Whether this contribution actually has connections to tail effects would be an intriguing topic to study.

Other physical effects that will only enter at higher perturbation orders are the effects from nonlinear radiation eikonal. 
The simplest example is the LO quadratic-in-graviton-field radiation eikonal $\chi_{(2)}^{H^2}$ corresponding to the 6-point tree amplitude, the latter argued to be irrelevant for classical physics~\cite{Cristofoli:2021jas}. 
Counting powers of $G$, the LO nonlinear radiation eikonal is expected to contribute at sufficiently high orders in the PM expansion; to the NNLO waveform (of order $\CO (G^{7/2})$) and the 5PM impulse, where the latter is currently state-of-the-art~\cite{Driesse:2024xad,Bern:2024adl,Driesse:2024feo}. 
If the LO nonlinear radiation eikonal $\chi_{(2)}^{H^2}$ indeed contributes to PM observables, then this would serve as a loophole to the claims made in Ref.~\cite{Cristofoli:2021jas}.

As a final note, it is remarked that the proposed radiation eikonal can be viewed as a classical limit of eikonal approximation applied to multiparticle scattering. 
To the best of author's knowledge, the eikonal approximation has mostly been applied to $2 \to 2$ scattering or its multiparticle deformations where particle production is viewed as ``soft'' corrections to the ``hard'' $2 \to 2$ scattering, and its application to general multiparticle kinematics has not been considered in the literature. 
What ``eikonal approximation of multiparticle scattering in general kinematics'' means in full quantum field theory is left as a food for thought to the reader.


\begin{acknowledgments}
\section{Acknowledgments}
The author would like to thank Carlo Heissenberg, Gustav Jakobsen, Joon-Hwi Kim, Sangmin Lee, Gustav Mogull, Nathan Moynihan, Julio Parra-Martinez, Jan Plefka, Trevor Scheopner, Jan Steinhoff, and Mao Zeng for insightful discussions.

\end{acknowledgments}

\bibliography{Manuscript.bib}

\begin{thebibliography}{75}%
\makeatletter
\providecommand \@ifxundefined [1]{%
 \@ifx{#1\undefined}
}%
\providecommand \@ifnum [1]{%
 \ifnum #1\expandafter \@firstoftwo
 \else \expandafter \@secondoftwo
 \fi
}%
\providecommand \@ifx [1]{%
 \ifx #1\expandafter \@firstoftwo
 \else \expandafter \@secondoftwo
 \fi
}%
\providecommand \natexlab [1]{#1}%
\providecommand \enquote  [1]{``#1''}%
\providecommand \bibnamefont  [1]{#1}%
\providecommand \bibfnamefont [1]{#1}%
\providecommand \citenamefont [1]{#1}%
\providecommand \href@noop [0]{\@secondoftwo}%
\providecommand \href [0]{\begingroup \@sanitize@url \@href}%
\providecommand \@href[1]{\@@startlink{#1}\@@href}%
\providecommand \@@href[1]{\endgroup#1\@@endlink}%
\providecommand \@sanitize@url [0]{\catcode `\\12\catcode `\$12\catcode
  `\&12\catcode `\#12\catcode `\^12\catcode `\_12\catcode `\%12\relax}%
\providecommand \@@startlink[1]{}%
\providecommand \@@endlink[0]{}%
\providecommand \url  [0]{\begingroup\@sanitize@url \@url }%
\providecommand \@url [1]{\endgroup\@href {#1}{\urlprefix }}%
\providecommand \urlprefix  [0]{URL }%
\providecommand \Eprint [0]{\href }%
\providecommand \doibase [0]{https://doi.org/}%
\providecommand \selectlanguage [0]{\@gobble}%
\providecommand \bibinfo  [0]{\@secondoftwo}%
\providecommand \bibfield  [0]{\@secondoftwo}%
\providecommand \translation [1]{[#1]}%
\providecommand \BibitemOpen [0]{}%
\providecommand \bibitemStop [0]{}%
\providecommand \bibitemNoStop [0]{.\EOS\space}%
\providecommand \EOS [0]{\spacefactor3000\relax}%
\providecommand \BibitemShut  [1]{\csname bibitem#1\endcsname}%
\let\auto@bib@innerbib\@empty
\bibitem [{\citenamefont {Abbott}\ \emph {et~al.}(2016)\citenamefont {Abbott}
  \emph {et~al.}}]{LIGOScientific:2016aoc}%
  \BibitemOpen
  \bibfield  {author} {\bibinfo {author} {\bibfnamefont {B.~P.}\ \bibnamefont
  {Abbott}} \emph {et~al.} (\bibinfo {collaboration} {LIGO Scientific,
  Virgo}),\ }\href {https://doi.org/10.1103/PhysRevLett.116.061102} {\bibfield
  {journal} {\bibinfo  {journal} {Phys. Rev. Lett.}\ }\textbf {\bibinfo
  {volume} {116}},\ \bibinfo {pages} {061102} (\bibinfo {year} {2016})},\
  \Eprint {https://arxiv.org/abs/1602.03837} {arXiv:1602.03837 [gr-qc]}
  \BibitemShut {NoStop}%
\bibitem [{\citenamefont {Damour}(2016)}]{Damour:2016gwp}%
  \BibitemOpen
  \bibfield  {author} {\bibinfo {author} {\bibfnamefont {T.}~\bibnamefont
  {Damour}},\ }\href {https://doi.org/10.1103/PhysRevD.94.104015} {\bibfield
  {journal} {\bibinfo  {journal} {Phys. Rev. D}\ }\textbf {\bibinfo {volume}
  {94}},\ \bibinfo {pages} {104015} (\bibinfo {year} {2016})},\ \Eprint
  {https://arxiv.org/abs/1609.00354} {arXiv:1609.00354 [gr-qc]} \BibitemShut
  {NoStop}%
\bibitem [{\citenamefont {Damour}(2018)}]{Damour:2017zjx}%
  \BibitemOpen
  \bibfield  {author} {\bibinfo {author} {\bibfnamefont {T.}~\bibnamefont
  {Damour}},\ }\href {https://doi.org/10.1103/PhysRevD.97.044038} {\bibfield
  {journal} {\bibinfo  {journal} {Phys. Rev. D}\ }\textbf {\bibinfo {volume}
  {97}},\ \bibinfo {pages} {044038} (\bibinfo {year} {2018})},\ \Eprint
  {https://arxiv.org/abs/1710.10599} {arXiv:1710.10599 [gr-qc]} \BibitemShut
  {NoStop}%
\bibitem [{\citenamefont {Westpfahl}\ and\ \citenamefont
  {Goller}(1979)}]{Westpfahl:1979gu}%
  \BibitemOpen
  \bibfield  {author} {\bibinfo {author} {\bibfnamefont {K.}~\bibnamefont
  {Westpfahl}}\ and\ \bibinfo {author} {\bibfnamefont {M.}~\bibnamefont
  {Goller}},\ }\href {https://doi.org/10.1007/BF02817047} {\bibfield  {journal}
  {\bibinfo  {journal} {Lett. Nuovo Cim.}\ }\textbf {\bibinfo {volume} {26}},\
  \bibinfo {pages} {573} (\bibinfo {year} {1979})}\BibitemShut {NoStop}%
\bibitem [{\citenamefont {Bern}\ \emph
  {et~al.}(2019{\natexlab{a}})\citenamefont {Bern}, \citenamefont {Cheung},
  \citenamefont {Roiban}, \citenamefont {Shen}, \citenamefont {Solon},\ and\
  \citenamefont {Zeng}}]{Bern:2019nnu}%
  \BibitemOpen
  \bibfield  {author} {\bibinfo {author} {\bibfnamefont {Z.}~\bibnamefont
  {Bern}}, \bibinfo {author} {\bibfnamefont {C.}~\bibnamefont {Cheung}},
  \bibinfo {author} {\bibfnamefont {R.}~\bibnamefont {Roiban}}, \bibinfo
  {author} {\bibfnamefont {C.-H.}\ \bibnamefont {Shen}}, \bibinfo {author}
  {\bibfnamefont {M.~P.}\ \bibnamefont {Solon}},\ and\ \bibinfo {author}
  {\bibfnamefont {M.}~\bibnamefont {Zeng}},\ }\href
  {https://doi.org/10.1103/PhysRevLett.122.201603} {\bibfield  {journal}
  {\bibinfo  {journal} {Phys. Rev. Lett.}\ }\textbf {\bibinfo {volume} {122}},\
  \bibinfo {pages} {201603} (\bibinfo {year} {2019}{\natexlab{a}})},\ \Eprint
  {https://arxiv.org/abs/1901.04424} {arXiv:1901.04424 [hep-th]} \BibitemShut
  {NoStop}%
\bibitem [{\citenamefont {Bern}\ \emph
  {et~al.}(2019{\natexlab{b}})\citenamefont {Bern}, \citenamefont {Cheung},
  \citenamefont {Roiban}, \citenamefont {Shen}, \citenamefont {Solon},\ and\
  \citenamefont {Zeng}}]{Bern:2019crd}%
  \BibitemOpen
  \bibfield  {author} {\bibinfo {author} {\bibfnamefont {Z.}~\bibnamefont
  {Bern}}, \bibinfo {author} {\bibfnamefont {C.}~\bibnamefont {Cheung}},
  \bibinfo {author} {\bibfnamefont {R.}~\bibnamefont {Roiban}}, \bibinfo
  {author} {\bibfnamefont {C.-H.}\ \bibnamefont {Shen}}, \bibinfo {author}
  {\bibfnamefont {M.~P.}\ \bibnamefont {Solon}},\ and\ \bibinfo {author}
  {\bibfnamefont {M.}~\bibnamefont {Zeng}},\ }\href
  {https://doi.org/10.1007/JHEP10(2019)206} {\bibfield  {journal} {\bibinfo
  {journal} {JHEP}\ }\textbf {\bibinfo {volume} {10}},\ \bibinfo {pages}
  {206}},\ \Eprint {https://arxiv.org/abs/1908.01493} {arXiv:1908.01493
  [hep-th]} \BibitemShut {NoStop}%
\bibitem [{\citenamefont {K\"alin}\ and\ \citenamefont
  {Porto}(2020)}]{Kalin:2020mvi}%
  \BibitemOpen
  \bibfield  {author} {\bibinfo {author} {\bibfnamefont {G.}~\bibnamefont
  {K\"alin}}\ and\ \bibinfo {author} {\bibfnamefont {R.~A.}\ \bibnamefont
  {Porto}},\ }\href {https://doi.org/10.1007/JHEP11(2020)106} {\bibfield
  {journal} {\bibinfo  {journal} {JHEP}\ }\textbf {\bibinfo {volume} {11}},\
  \bibinfo {pages} {106}},\ \Eprint {https://arxiv.org/abs/2006.01184}
  {arXiv:2006.01184 [hep-th]} \BibitemShut {NoStop}%
\bibitem [{\citenamefont {K\"alin}\ \emph {et~al.}(2020)\citenamefont
  {K\"alin}, \citenamefont {Liu},\ and\ \citenamefont {Porto}}]{Kalin:2020fhe}%
  \BibitemOpen
  \bibfield  {author} {\bibinfo {author} {\bibfnamefont {G.}~\bibnamefont
  {K\"alin}}, \bibinfo {author} {\bibfnamefont {Z.}~\bibnamefont {Liu}},\ and\
  \bibinfo {author} {\bibfnamefont {R.~A.}\ \bibnamefont {Porto}},\ }\href
  {https://doi.org/10.1103/PhysRevLett.125.261103} {\bibfield  {journal}
  {\bibinfo  {journal} {Phys. Rev. Lett.}\ }\textbf {\bibinfo {volume} {125}},\
  \bibinfo {pages} {261103} (\bibinfo {year} {2020})},\ \Eprint
  {https://arxiv.org/abs/2007.04977} {arXiv:2007.04977 [hep-th]} \BibitemShut
  {NoStop}%
\bibitem [{\citenamefont {Bern}\ \emph
  {et~al.}(2021{\natexlab{a}})\citenamefont {Bern}, \citenamefont
  {Parra-Martinez}, \citenamefont {Roiban}, \citenamefont {Ruf}, \citenamefont
  {Shen}, \citenamefont {Solon},\ and\ \citenamefont {Zeng}}]{Bern:2021dqo}%
  \BibitemOpen
  \bibfield  {author} {\bibinfo {author} {\bibfnamefont {Z.}~\bibnamefont
  {Bern}}, \bibinfo {author} {\bibfnamefont {J.}~\bibnamefont
  {Parra-Martinez}}, \bibinfo {author} {\bibfnamefont {R.}~\bibnamefont
  {Roiban}}, \bibinfo {author} {\bibfnamefont {M.~S.}\ \bibnamefont {Ruf}},
  \bibinfo {author} {\bibfnamefont {C.-H.}\ \bibnamefont {Shen}}, \bibinfo
  {author} {\bibfnamefont {M.~P.}\ \bibnamefont {Solon}},\ and\ \bibinfo
  {author} {\bibfnamefont {M.}~\bibnamefont {Zeng}},\ }\href
  {https://doi.org/10.1103/PhysRevLett.126.171601} {\bibfield  {journal}
  {\bibinfo  {journal} {Phys. Rev. Lett.}\ }\textbf {\bibinfo {volume} {126}},\
  \bibinfo {pages} {171601} (\bibinfo {year} {2021}{\natexlab{a}})},\ \Eprint
  {https://arxiv.org/abs/2101.07254} {arXiv:2101.07254 [hep-th]} \BibitemShut
  {NoStop}%
\bibitem [{\citenamefont {Dlapa}\ \emph {et~al.}(2022)\citenamefont {Dlapa},
  \citenamefont {K\"alin}, \citenamefont {Liu},\ and\ \citenamefont
  {Porto}}]{Dlapa:2021npj}%
  \BibitemOpen
  \bibfield  {author} {\bibinfo {author} {\bibfnamefont {C.}~\bibnamefont
  {Dlapa}}, \bibinfo {author} {\bibfnamefont {G.}~\bibnamefont {K\"alin}},
  \bibinfo {author} {\bibfnamefont {Z.}~\bibnamefont {Liu}},\ and\ \bibinfo
  {author} {\bibfnamefont {R.~A.}\ \bibnamefont {Porto}},\ }\href
  {https://doi.org/10.1016/j.physletb.2022.137203} {\bibfield  {journal}
  {\bibinfo  {journal} {Phys. Lett. B}\ }\textbf {\bibinfo {volume} {831}},\
  \bibinfo {pages} {137203} (\bibinfo {year} {2022})},\ \Eprint
  {https://arxiv.org/abs/2106.08276} {arXiv:2106.08276 [hep-th]} \BibitemShut
  {NoStop}%
\bibitem [{\citenamefont {Jakobsen}\ \emph {et~al.}(2022)\citenamefont
  {Jakobsen}, \citenamefont {Mogull}, \citenamefont {Plefka},\ and\
  \citenamefont {Sauer}}]{Jakobsen:2022psy}%
  \BibitemOpen
  \bibfield  {author} {\bibinfo {author} {\bibfnamefont {G.~U.}\ \bibnamefont
  {Jakobsen}}, \bibinfo {author} {\bibfnamefont {G.}~\bibnamefont {Mogull}},
  \bibinfo {author} {\bibfnamefont {J.}~\bibnamefont {Plefka}},\ and\ \bibinfo
  {author} {\bibfnamefont {B.}~\bibnamefont {Sauer}},\ }\href
  {https://doi.org/10.1007/JHEP10(2022)128} {\bibfield  {journal} {\bibinfo
  {journal} {JHEP}\ }\textbf {\bibinfo {volume} {10}},\ \bibinfo {pages}
  {128}},\ \Eprint {https://arxiv.org/abs/2207.00569} {arXiv:2207.00569
  [hep-th]} \BibitemShut {NoStop}%
\bibitem [{\citenamefont {K\"alin}\ \emph {et~al.}(2023)\citenamefont
  {K\"alin}, \citenamefont {Neef},\ and\ \citenamefont
  {Porto}}]{Kalin:2022hph}%
  \BibitemOpen
  \bibfield  {author} {\bibinfo {author} {\bibfnamefont {G.}~\bibnamefont
  {K\"alin}}, \bibinfo {author} {\bibfnamefont {J.}~\bibnamefont {Neef}},\ and\
  \bibinfo {author} {\bibfnamefont {R.~A.}\ \bibnamefont {Porto}},\ }\href
  {https://doi.org/10.1007/JHEP01(2023)140} {\bibfield  {journal} {\bibinfo
  {journal} {JHEP}\ }\textbf {\bibinfo {volume} {01}},\ \bibinfo {pages}
  {140}},\ \Eprint {https://arxiv.org/abs/2207.00580} {arXiv:2207.00580
  [hep-th]} \BibitemShut {NoStop}%
\bibitem [{\citenamefont {Dlapa}\ \emph {et~al.}(2023)\citenamefont {Dlapa},
  \citenamefont {K\"alin}, \citenamefont {Liu}, \citenamefont {Neef},\ and\
  \citenamefont {Porto}}]{Dlapa:2022lmu}%
  \BibitemOpen
  \bibfield  {author} {\bibinfo {author} {\bibfnamefont {C.}~\bibnamefont
  {Dlapa}}, \bibinfo {author} {\bibfnamefont {G.}~\bibnamefont {K\"alin}},
  \bibinfo {author} {\bibfnamefont {Z.}~\bibnamefont {Liu}}, \bibinfo {author}
  {\bibfnamefont {J.}~\bibnamefont {Neef}},\ and\ \bibinfo {author}
  {\bibfnamefont {R.~A.}\ \bibnamefont {Porto}},\ }\href
  {https://doi.org/10.1103/PhysRevLett.130.101401} {\bibfield  {journal}
  {\bibinfo  {journal} {Phys. Rev. Lett.}\ }\textbf {\bibinfo {volume} {130}},\
  \bibinfo {pages} {101401} (\bibinfo {year} {2023})},\ \Eprint
  {https://arxiv.org/abs/2210.05541} {arXiv:2210.05541 [hep-th]} \BibitemShut
  {NoStop}%
\bibitem [{\citenamefont {Dlapa}\ \emph {et~al.}(2024)\citenamefont {Dlapa},
  \citenamefont {K\"alin}, \citenamefont {Liu},\ and\ \citenamefont
  {Porto}}]{Dlapa:2024cje}%
  \BibitemOpen
  \bibfield  {author} {\bibinfo {author} {\bibfnamefont {C.}~\bibnamefont
  {Dlapa}}, \bibinfo {author} {\bibfnamefont {G.}~\bibnamefont {K\"alin}},
  \bibinfo {author} {\bibfnamefont {Z.}~\bibnamefont {Liu}},\ and\ \bibinfo
  {author} {\bibfnamefont {R.~A.}\ \bibnamefont {Porto}},\ }\href
  {https://doi.org/10.1103/PhysRevLett.132.221401} {\bibfield  {journal}
  {\bibinfo  {journal} {Phys. Rev. Lett.}\ }\textbf {\bibinfo {volume} {132}},\
  \bibinfo {pages} {221401} (\bibinfo {year} {2024})},\ \Eprint
  {https://arxiv.org/abs/2403.04853} {arXiv:2403.04853 [hep-th]} \BibitemShut
  {NoStop}%
\bibitem [{\citenamefont {Driesse}\ \emph
  {et~al.}(2024{\natexlab{a}})\citenamefont {Driesse}, \citenamefont
  {Jakobsen}, \citenamefont {Mogull}, \citenamefont {Plefka}, \citenamefont
  {Sauer},\ and\ \citenamefont {Usovitsch}}]{Driesse:2024xad}%
  \BibitemOpen
  \bibfield  {author} {\bibinfo {author} {\bibfnamefont {M.}~\bibnamefont
  {Driesse}}, \bibinfo {author} {\bibfnamefont {G.~U.}\ \bibnamefont
  {Jakobsen}}, \bibinfo {author} {\bibfnamefont {G.}~\bibnamefont {Mogull}},
  \bibinfo {author} {\bibfnamefont {J.}~\bibnamefont {Plefka}}, \bibinfo
  {author} {\bibfnamefont {B.}~\bibnamefont {Sauer}},\ and\ \bibinfo {author}
  {\bibfnamefont {J.}~\bibnamefont {Usovitsch}},\ }\href
  {https://doi.org/10.1103/PhysRevLett.132.241402} {\bibfield  {journal}
  {\bibinfo  {journal} {Phys. Rev. Lett.}\ }\textbf {\bibinfo {volume} {132}},\
  \bibinfo {pages} {241402} (\bibinfo {year} {2024}{\natexlab{a}})},\ \Eprint
  {https://arxiv.org/abs/2403.07781} {arXiv:2403.07781 [hep-th]} \BibitemShut
  {NoStop}%
\bibitem [{\citenamefont {Bern}\ \emph {et~al.}(2024)\citenamefont {Bern},
  \citenamefont {Herrmann}, \citenamefont {Roiban}, \citenamefont {Ruf},
  \citenamefont {Smirnov}, \citenamefont {Smirnov},\ and\ \citenamefont
  {Zeng}}]{Bern:2024adl}%
  \BibitemOpen
  \bibfield  {author} {\bibinfo {author} {\bibfnamefont {Z.}~\bibnamefont
  {Bern}}, \bibinfo {author} {\bibfnamefont {E.}~\bibnamefont {Herrmann}},
  \bibinfo {author} {\bibfnamefont {R.}~\bibnamefont {Roiban}}, \bibinfo
  {author} {\bibfnamefont {M.~S.}\ \bibnamefont {Ruf}}, \bibinfo {author}
  {\bibfnamefont {A.~V.}\ \bibnamefont {Smirnov}}, \bibinfo {author}
  {\bibfnamefont {V.~A.}\ \bibnamefont {Smirnov}},\ and\ \bibinfo {author}
  {\bibfnamefont {M.}~\bibnamefont {Zeng}},\ }\href
  {https://doi.org/10.1007/JHEP10(2024)023} {\bibfield  {journal} {\bibinfo
  {journal} {JHEP}\ }\textbf {\bibinfo {volume} {10}},\ \bibinfo {pages}
  {023}},\ \Eprint {https://arxiv.org/abs/2406.01554} {arXiv:2406.01554
  [hep-th]} \BibitemShut {NoStop}%
\bibitem [{\citenamefont {Driesse}\ \emph
  {et~al.}(2024{\natexlab{b}})\citenamefont {Driesse}, \citenamefont
  {Jakobsen}, \citenamefont {Klemm}, \citenamefont {Mogull}, \citenamefont
  {Nega}, \citenamefont {Plefka}, \citenamefont {Sauer},\ and\ \citenamefont
  {Usovitsch}}]{Driesse:2024feo}%
  \BibitemOpen
  \bibfield  {author} {\bibinfo {author} {\bibfnamefont {M.}~\bibnamefont
  {Driesse}}, \bibinfo {author} {\bibfnamefont {G.~U.}\ \bibnamefont
  {Jakobsen}}, \bibinfo {author} {\bibfnamefont {A.}~\bibnamefont {Klemm}},
  \bibinfo {author} {\bibfnamefont {G.}~\bibnamefont {Mogull}}, \bibinfo
  {author} {\bibfnamefont {C.}~\bibnamefont {Nega}}, \bibinfo {author}
  {\bibfnamefont {J.}~\bibnamefont {Plefka}}, \bibinfo {author} {\bibfnamefont
  {B.}~\bibnamefont {Sauer}},\ and\ \bibinfo {author} {\bibfnamefont
  {J.}~\bibnamefont {Usovitsch}},\ }\href@noop {} {\  (\bibinfo {year}
  {2024}{\natexlab{b}})},\ \Eprint {https://arxiv.org/abs/2411.11846}
  {arXiv:2411.11846 [hep-th]} \BibitemShut {NoStop}%
\bibitem [{\citenamefont {Bjerrum-Bohr}\ \emph {et~al.}(2022)\citenamefont
  {Bjerrum-Bohr}, \citenamefont {Damgaard}, \citenamefont {Plante},\ and\
  \citenamefont {Vanhove}}]{Bjerrum-Bohr:2022blt}%
  \BibitemOpen
  \bibfield  {author} {\bibinfo {author} {\bibfnamefont {N.~E.~J.}\
  \bibnamefont {Bjerrum-Bohr}}, \bibinfo {author} {\bibfnamefont {P.~H.}\
  \bibnamefont {Damgaard}}, \bibinfo {author} {\bibfnamefont {L.}~\bibnamefont
  {Plante}},\ and\ \bibinfo {author} {\bibfnamefont {P.}~\bibnamefont
  {Vanhove}},\ }\href {https://doi.org/10.1088/1751-8121/ac7a78} {\bibfield
  {journal} {\bibinfo  {journal} {J. Phys. A}\ }\textbf {\bibinfo {volume}
  {55}},\ \bibinfo {pages} {443014} (\bibinfo {year} {2022})},\ \Eprint
  {https://arxiv.org/abs/2203.13024} {arXiv:2203.13024 [hep-th]} \BibitemShut
  {NoStop}%
\bibitem [{\citenamefont {Kosower}\ \emph {et~al.}(2022)\citenamefont
  {Kosower}, \citenamefont {Monteiro},\ and\ \citenamefont
  {O'Connell}}]{Kosower:2022yvp}%
  \BibitemOpen
  \bibfield  {author} {\bibinfo {author} {\bibfnamefont {D.~A.}\ \bibnamefont
  {Kosower}}, \bibinfo {author} {\bibfnamefont {R.}~\bibnamefont {Monteiro}},\
  and\ \bibinfo {author} {\bibfnamefont {D.}~\bibnamefont {O'Connell}},\ }\href
  {https://doi.org/10.1088/1751-8121/ac8846} {\bibfield  {journal} {\bibinfo
  {journal} {J. Phys. A}\ }\textbf {\bibinfo {volume} {55}},\ \bibinfo {pages}
  {443015} (\bibinfo {year} {2022})},\ \Eprint
  {https://arxiv.org/abs/2203.13025} {arXiv:2203.13025 [hep-th]} \BibitemShut
  {NoStop}%
\bibitem [{\citenamefont {Buonanno}\ \emph {et~al.}(2022)\citenamefont
  {Buonanno}, \citenamefont {Khalil}, \citenamefont {O'Connell}, \citenamefont
  {Roiban}, \citenamefont {Solon},\ and\ \citenamefont
  {Zeng}}]{Buonanno:2022pgc}%
  \BibitemOpen
  \bibfield  {author} {\bibinfo {author} {\bibfnamefont {A.}~\bibnamefont
  {Buonanno}}, \bibinfo {author} {\bibfnamefont {M.}~\bibnamefont {Khalil}},
  \bibinfo {author} {\bibfnamefont {D.}~\bibnamefont {O'Connell}}, \bibinfo
  {author} {\bibfnamefont {R.}~\bibnamefont {Roiban}}, \bibinfo {author}
  {\bibfnamefont {M.~P.}\ \bibnamefont {Solon}},\ and\ \bibinfo {author}
  {\bibfnamefont {M.}~\bibnamefont {Zeng}},\ }in\ \href@noop {} {\emph
  {\bibinfo {booktitle} {{Snowmass 2021}}}}\ (\bibinfo {year} {2022})\ \Eprint
  {https://arxiv.org/abs/2204.05194} {arXiv:2204.05194 [hep-th]} \BibitemShut
  {NoStop}%
\bibitem [{Note1()}]{Note1}%
  \BibitemOpen
  \bibinfo {note} {The existence of the eikonal is proven for conservative
  dynamics, and this work addresses inclusion of radiative effects. Note that
  the eikonal may not be single-valued. Well-definedness of its derivatives
  suffices, since the eikonal itself is not an observable quantity; only though
  its difference the eikonal represents observable quantities, even for the
  quantities that cannot be computed from the scattering generator equation
  [Eq.\protect \textup {\hbox {\mathsurround \z@ \protect \normalfont
  (\ignorespaces \ref {eq:scgen_eq}\unskip \@@italiccorr )}}] such as the time
  delay~\cite {Shapiro:1964uw,Camanho:2014apa} and polarisation rotation~\cite
  {Chen:2022clh,Kim:2022iub}.}\BibitemShut {Stop}%
\bibitem [{\citenamefont {Buonanno}\ and\ \citenamefont
  {Damour}(1999)}]{Buonanno:1998gg}%
  \BibitemOpen
  \bibfield  {author} {\bibinfo {author} {\bibfnamefont {A.}~\bibnamefont
  {Buonanno}}\ and\ \bibinfo {author} {\bibfnamefont {T.}~\bibnamefont
  {Damour}},\ }\href {https://doi.org/10.1103/PhysRevD.59.084006} {\bibfield
  {journal} {\bibinfo  {journal} {Phys. Rev. D}\ }\textbf {\bibinfo {volume}
  {59}},\ \bibinfo {pages} {084006} (\bibinfo {year} {1999})},\ \Eprint
  {https://arxiv.org/abs/gr-qc/9811091} {arXiv:gr-qc/9811091} \BibitemShut
  {NoStop}%
\bibitem [{\citenamefont {Pompili}\ \emph {et~al.}(2023)\citenamefont {Pompili}
  \emph {et~al.}}]{Pompili:2023tna}%
  \BibitemOpen
  \bibfield  {author} {\bibinfo {author} {\bibfnamefont {L.}~\bibnamefont
  {Pompili}} \emph {et~al.},\ }\href
  {https://doi.org/10.1103/PhysRevD.108.124035} {\bibfield  {journal} {\bibinfo
   {journal} {Phys. Rev. D}\ }\textbf {\bibinfo {volume} {108}},\ \bibinfo
  {pages} {124035} (\bibinfo {year} {2023})},\ \Eprint
  {https://arxiv.org/abs/2303.18039} {arXiv:2303.18039 [gr-qc]} \BibitemShut
  {NoStop}%
\bibitem [{\citenamefont {Buonanno}\ \emph {et~al.}(2024)\citenamefont
  {Buonanno}, \citenamefont {Mogull}, \citenamefont {Patil},\ and\
  \citenamefont {Pompili}}]{Buonanno:2024byg}%
  \BibitemOpen
  \bibfield  {author} {\bibinfo {author} {\bibfnamefont {A.}~\bibnamefont
  {Buonanno}}, \bibinfo {author} {\bibfnamefont {G.}~\bibnamefont {Mogull}},
  \bibinfo {author} {\bibfnamefont {R.}~\bibnamefont {Patil}},\ and\ \bibinfo
  {author} {\bibfnamefont {L.}~\bibnamefont {Pompili}},\ }\href
  {https://doi.org/10.1103/PhysRevLett.133.211402} {\bibfield  {journal}
  {\bibinfo  {journal} {Phys. Rev. Lett.}\ }\textbf {\bibinfo {volume} {133}},\
  \bibinfo {pages} {211402} (\bibinfo {year} {2024})},\ \Eprint
  {https://arxiv.org/abs/2405.19181} {arXiv:2405.19181 [gr-qc]} \BibitemShut
  {NoStop}%
\bibitem [{\citenamefont {Kim}\ \emph {et~al.}(2024{\natexlab{a}})\citenamefont
  {Kim}, \citenamefont {Kim},\ and\ \citenamefont {Lee}}]{Kim:2024grz}%
  \BibitemOpen
  \bibfield  {author} {\bibinfo {author} {\bibfnamefont {J.-H.}\ \bibnamefont
  {Kim}}, \bibinfo {author} {\bibfnamefont {J.-W.}\ \bibnamefont {Kim}},\ and\
  \bibinfo {author} {\bibfnamefont {S.}~\bibnamefont {Lee}},\ }\href
  {https://doi.org/10.1007/JHEP08(2024)080} {\bibfield  {journal} {\bibinfo
  {journal} {JHEP}\ }\textbf {\bibinfo {volume} {08}},\ \bibinfo {pages}
  {080}},\ \Eprint {https://arxiv.org/abs/2405.17056} {arXiv:2405.17056
  [hep-th]} \BibitemShut {NoStop}%
\bibitem [{\citenamefont {Chen}\ \emph {et~al.}(2024)\citenamefont {Chen},
  \citenamefont {Kim},\ and\ \citenamefont {Wang}}]{Chen:2024bpf}%
  \BibitemOpen
  \bibfield  {author} {\bibinfo {author} {\bibfnamefont {G.}~\bibnamefont
  {Chen}}, \bibinfo {author} {\bibfnamefont {J.-W.}\ \bibnamefont {Kim}},\ and\
  \bibinfo {author} {\bibfnamefont {T.}~\bibnamefont {Wang}},\ }\href@noop {}
  {\  (\bibinfo {year} {2024})},\ \Eprint {https://arxiv.org/abs/2406.17658}
  {arXiv:2406.17658 [hep-th]} \BibitemShut {NoStop}%
\bibitem [{\citenamefont {Kim}(2025)}]{Kim:2024zsf}%
  \BibitemOpen
  \bibfield  {author} {\bibinfo {author} {\bibfnamefont {J.-W.}\ \bibnamefont
  {Kim}},\ }\href {https://doi.org/10.22323/1.476.0790} {\bibfield  {journal}
  {\bibinfo  {journal} {PoS}\ }\textbf {\bibinfo {volume} {ICHEP2024}},\
  \bibinfo {pages} {790} (\bibinfo {year} {2025})}\BibitemShut {NoStop}%
\bibitem [{\citenamefont {Kim}\ \emph {et~al.}(2024{\natexlab{b}})\citenamefont
  {Kim}, \citenamefont {Kim}, \citenamefont {Kim},\ and\ \citenamefont
  {Lee}}]{Kim:2024svw}%
  \BibitemOpen
  \bibfield  {author} {\bibinfo {author} {\bibfnamefont {J.-H.}\ \bibnamefont
  {Kim}}, \bibinfo {author} {\bibfnamefont {J.-W.}\ \bibnamefont {Kim}},
  \bibinfo {author} {\bibfnamefont {S.}~\bibnamefont {Kim}},\ and\ \bibinfo
  {author} {\bibfnamefont {S.}~\bibnamefont {Lee}},\ }\href@noop {} {\
  (\bibinfo {year} {2024}{\natexlab{b}})},\ \Eprint
  {https://arxiv.org/abs/2410.22988} {arXiv:2410.22988 [hep-th]} \BibitemShut
  {NoStop}%
\bibitem [{\citenamefont {Gonzo}\ and\ \citenamefont
  {Shi}(2024)}]{Gonzo:2024zxo}%
  \BibitemOpen
  \bibfield  {author} {\bibinfo {author} {\bibfnamefont {R.}~\bibnamefont
  {Gonzo}}\ and\ \bibinfo {author} {\bibfnamefont {C.}~\bibnamefont {Shi}},\
  }\href {https://doi.org/10.1103/PhysRevLett.133.221401} {\bibfield  {journal}
  {\bibinfo  {journal} {Phys. Rev. Lett.}\ }\textbf {\bibinfo {volume} {133}},\
  \bibinfo {pages} {221401} (\bibinfo {year} {2024})},\ \Eprint
  {https://arxiv.org/abs/2405.09687} {arXiv:2405.09687 [hep-th]} \BibitemShut
  {NoStop}%
\bibitem [{\citenamefont {Kosower}\ \emph {et~al.}(2019)\citenamefont
  {Kosower}, \citenamefont {Maybee},\ and\ \citenamefont
  {O'Connell}}]{Kosower:2018adc}%
  \BibitemOpen
  \bibfield  {author} {\bibinfo {author} {\bibfnamefont {D.~A.}\ \bibnamefont
  {Kosower}}, \bibinfo {author} {\bibfnamefont {B.}~\bibnamefont {Maybee}},\
  and\ \bibinfo {author} {\bibfnamefont {D.}~\bibnamefont {O'Connell}},\ }\href
  {https://doi.org/10.1007/JHEP02(2019)137} {\bibfield  {journal} {\bibinfo
  {journal} {JHEP}\ }\textbf {\bibinfo {volume} {02}},\ \bibinfo {pages}
  {137}},\ \Eprint {https://arxiv.org/abs/1811.10950} {arXiv:1811.10950
  [hep-th]} \BibitemShut {NoStop}%
\bibitem [{\citenamefont {Lehmann}\ \emph {et~al.}(1957)\citenamefont
  {Lehmann}, \citenamefont {Symanzik},\ and\ \citenamefont
  {Zimmermann}}]{Lehmann:1957zz}%
  \BibitemOpen
  \bibfield  {author} {\bibinfo {author} {\bibfnamefont {H.}~\bibnamefont
  {Lehmann}}, \bibinfo {author} {\bibfnamefont {K.}~\bibnamefont {Symanzik}},\
  and\ \bibinfo {author} {\bibfnamefont {W.}~\bibnamefont {Zimmermann}},\
  }\href {https://doi.org/10.1007/BF02832508} {\bibfield  {journal} {\bibinfo
  {journal} {Nuovo Cim.}\ }\textbf {\bibinfo {volume} {6}},\ \bibinfo {pages}
  {319} (\bibinfo {year} {1957})}\BibitemShut {NoStop}%
\bibitem [{\citenamefont {Damgaard}\ \emph {et~al.}(2021)\citenamefont
  {Damgaard}, \citenamefont {Plante},\ and\ \citenamefont
  {Vanhove}}]{Damgaard:2021ipf}%
  \BibitemOpen
  \bibfield  {author} {\bibinfo {author} {\bibfnamefont {P.~H.}\ \bibnamefont
  {Damgaard}}, \bibinfo {author} {\bibfnamefont {L.}~\bibnamefont {Plante}},\
  and\ \bibinfo {author} {\bibfnamefont {P.}~\bibnamefont {Vanhove}},\ }\href
  {https://doi.org/10.1007/JHEP11(2021)213} {\bibfield  {journal} {\bibinfo
  {journal} {JHEP}\ }\textbf {\bibinfo {volume} {11}},\ \bibinfo {pages}
  {213}},\ \Eprint {https://arxiv.org/abs/2107.12891} {arXiv:2107.12891
  [hep-th]} \BibitemShut {NoStop}%
\bibitem [{\citenamefont {Damgaard}\ \emph {et~al.}(2023)\citenamefont
  {Damgaard}, \citenamefont {Hansen}, \citenamefont {Plant\'e},\ and\
  \citenamefont {Vanhove}}]{Damgaard:2023ttc}%
  \BibitemOpen
  \bibfield  {author} {\bibinfo {author} {\bibfnamefont {P.~H.}\ \bibnamefont
  {Damgaard}}, \bibinfo {author} {\bibfnamefont {E.~R.}\ \bibnamefont
  {Hansen}}, \bibinfo {author} {\bibfnamefont {L.}~\bibnamefont {Plant\'e}},\
  and\ \bibinfo {author} {\bibfnamefont {P.}~\bibnamefont {Vanhove}},\ }\href
  {https://doi.org/10.1007/JHEP09(2023)183} {\bibfield  {journal} {\bibinfo
  {journal} {JHEP}\ }\textbf {\bibinfo {volume} {09}},\ \bibinfo {pages}
  {183}},\ \Eprint {https://arxiv.org/abs/2307.04746} {arXiv:2307.04746
  [hep-th]} \BibitemShut {NoStop}%
\bibitem [{\citenamefont {Di~Vecchia}\ \emph {et~al.}(2024)\citenamefont
  {Di~Vecchia}, \citenamefont {Heissenberg}, \citenamefont {Russo},\ and\
  \citenamefont {Veneziano}}]{DiVecchia:2023frv}%
  \BibitemOpen
  \bibfield  {author} {\bibinfo {author} {\bibfnamefont {P.}~\bibnamefont
  {Di~Vecchia}}, \bibinfo {author} {\bibfnamefont {C.}~\bibnamefont
  {Heissenberg}}, \bibinfo {author} {\bibfnamefont {R.}~\bibnamefont {Russo}},\
  and\ \bibinfo {author} {\bibfnamefont {G.}~\bibnamefont {Veneziano}},\ }\href
  {https://doi.org/10.1016/j.physrep.2024.06.002} {\bibfield  {journal}
  {\bibinfo  {journal} {Phys. Rept.}\ }\textbf {\bibinfo {volume} {1083}},\
  \bibinfo {pages} {1} (\bibinfo {year} {2024})},\ \Eprint
  {https://arxiv.org/abs/2306.16488} {arXiv:2306.16488 [hep-th]} \BibitemShut
  {NoStop}%
\bibitem [{\citenamefont {Georgoudis}\ \emph
  {et~al.}(2024{\natexlab{a}})\citenamefont {Georgoudis}, \citenamefont
  {Heissenberg},\ and\ \citenamefont {Russo}}]{Georgoudis:2023eke}%
  \BibitemOpen
  \bibfield  {author} {\bibinfo {author} {\bibfnamefont {A.}~\bibnamefont
  {Georgoudis}}, \bibinfo {author} {\bibfnamefont {C.}~\bibnamefont
  {Heissenberg}},\ and\ \bibinfo {author} {\bibfnamefont {R.}~\bibnamefont
  {Russo}},\ }\href {https://doi.org/10.1007/JHEP03(2024)089} {\bibfield
  {journal} {\bibinfo  {journal} {JHEP}\ }\textbf {\bibinfo {volume} {03}},\
  \bibinfo {pages} {089}},\ \Eprint {https://arxiv.org/abs/2312.07452}
  {arXiv:2312.07452 [hep-th]} \BibitemShut {NoStop}%
\bibitem [{\citenamefont {Luna}\ \emph {et~al.}(2024)\citenamefont {Luna},
  \citenamefont {Moynihan}, \citenamefont {O'Connell},\ and\ \citenamefont
  {Ross}}]{Luna:2023uwd}%
  \BibitemOpen
  \bibfield  {author} {\bibinfo {author} {\bibfnamefont {A.}~\bibnamefont
  {Luna}}, \bibinfo {author} {\bibfnamefont {N.}~\bibnamefont {Moynihan}},
  \bibinfo {author} {\bibfnamefont {D.}~\bibnamefont {O'Connell}},\ and\
  \bibinfo {author} {\bibfnamefont {A.}~\bibnamefont {Ross}},\ }\href
  {https://doi.org/10.1007/JHEP08(2024)045} {\bibfield  {journal} {\bibinfo
  {journal} {JHEP}\ }\textbf {\bibinfo {volume} {08}},\ \bibinfo {pages}
  {045}},\ \Eprint {https://arxiv.org/abs/2312.09960} {arXiv:2312.09960
  [hep-th]} \BibitemShut {NoStop}%
\bibitem [{\citenamefont {Di~Vecchia}\ \emph
  {et~al.}(2022{\natexlab{a}})\citenamefont {Di~Vecchia}, \citenamefont
  {Heissenberg}, \citenamefont {Russo},\ and\ \citenamefont
  {Veneziano}}]{DiVecchia:2022nna}%
  \BibitemOpen
  \bibfield  {author} {\bibinfo {author} {\bibfnamefont {P.}~\bibnamefont
  {Di~Vecchia}}, \bibinfo {author} {\bibfnamefont {C.}~\bibnamefont
  {Heissenberg}}, \bibinfo {author} {\bibfnamefont {R.}~\bibnamefont {Russo}},\
  and\ \bibinfo {author} {\bibfnamefont {G.}~\bibnamefont {Veneziano}},\ }\href
  {https://doi.org/10.1007/JHEP07(2022)039} {\bibfield  {journal} {\bibinfo
  {journal} {JHEP}\ }\textbf {\bibinfo {volume} {07}},\ \bibinfo {pages}
  {039}},\ \Eprint {https://arxiv.org/abs/2204.02378} {arXiv:2204.02378
  [hep-th]} \BibitemShut {NoStop}%
\bibitem [{\citenamefont {Di~Vecchia}\ \emph {et~al.}(2023)\citenamefont
  {Di~Vecchia}, \citenamefont {Heissenberg}, \citenamefont {Russo},\ and\
  \citenamefont {Veneziano}}]{DiVecchia:2022piu}%
  \BibitemOpen
  \bibfield  {author} {\bibinfo {author} {\bibfnamefont {P.}~\bibnamefont
  {Di~Vecchia}}, \bibinfo {author} {\bibfnamefont {C.}~\bibnamefont
  {Heissenberg}}, \bibinfo {author} {\bibfnamefont {R.}~\bibnamefont {Russo}},\
  and\ \bibinfo {author} {\bibfnamefont {G.}~\bibnamefont {Veneziano}},\ }\href
  {https://doi.org/10.1016/j.physletb.2023.138049} {\bibfield  {journal}
  {\bibinfo  {journal} {Phys. Lett. B}\ }\textbf {\bibinfo {volume} {843}},\
  \bibinfo {pages} {138049} (\bibinfo {year} {2023})},\ \Eprint
  {https://arxiv.org/abs/2210.12118} {arXiv:2210.12118 [hep-th]} \BibitemShut
  {NoStop}%
\bibitem [{\citenamefont {Aoude}\ \emph {et~al.}(2024)\citenamefont {Aoude},
  \citenamefont {Cristofoli}, \citenamefont {Elkhidir},\ and\ \citenamefont
  {Sergola}}]{Aoude:2024jxd}%
  \BibitemOpen
  \bibfield  {author} {\bibinfo {author} {\bibfnamefont {R.}~\bibnamefont
  {Aoude}}, \bibinfo {author} {\bibfnamefont {A.}~\bibnamefont {Cristofoli}},
  \bibinfo {author} {\bibfnamefont {A.}~\bibnamefont {Elkhidir}},\ and\
  \bibinfo {author} {\bibfnamefont {M.}~\bibnamefont {Sergola}},\ }\href@noop
  {} {\  (\bibinfo {year} {2024})},\ \Eprint {https://arxiv.org/abs/2411.02294}
  {arXiv:2411.02294 [hep-th]} \BibitemShut {NoStop}%
\bibitem [{\citenamefont {Mogull}\ \emph {et~al.}(2021)\citenamefont {Mogull},
  \citenamefont {Plefka},\ and\ \citenamefont {Steinhoff}}]{Mogull:2020sak}%
  \BibitemOpen
  \bibfield  {author} {\bibinfo {author} {\bibfnamefont {G.}~\bibnamefont
  {Mogull}}, \bibinfo {author} {\bibfnamefont {J.}~\bibnamefont {Plefka}},\
  and\ \bibinfo {author} {\bibfnamefont {J.}~\bibnamefont {Steinhoff}},\ }\href
  {https://doi.org/10.1007/JHEP02(2021)048} {\bibfield  {journal} {\bibinfo
  {journal} {JHEP}\ }\textbf {\bibinfo {volume} {02}},\ \bibinfo {pages}
  {048}},\ \Eprint {https://arxiv.org/abs/2010.02865} {arXiv:2010.02865
  [hep-th]} \BibitemShut {NoStop}%
\bibitem [{\citenamefont {{Mart{\'i}n-Garc{\'i}a, J. M.}}()}]{xAct}%
  \BibitemOpen
  \bibfield  {author} {\bibinfo {author} {\bibnamefont {{Mart{\'i}n-Garc{\'i}a,
  J. M.}}},\ }\href {{http://xact.es/}} {\bibinfo {title} {{xAct: Efficient
  tensor computer algebra for the Wolfram Language}}},\ \bibinfo {howpublished}
  {\url{http://xact.es/}}\BibitemShut {NoStop}%
\bibitem [{\citenamefont {Lee}(2012)}]{Lee:2012cn}%
  \BibitemOpen
  \bibfield  {author} {\bibinfo {author} {\bibfnamefont {R.~N.}\ \bibnamefont
  {Lee}},\ }\href@noop {} {\  (\bibinfo {year} {2012})},\ \Eprint
  {https://arxiv.org/abs/1212.2685} {arXiv:1212.2685 [hep-ph]} \BibitemShut
  {NoStop}%
\bibitem [{\citenamefont {Peierls}(1952)}]{Peierls:1952cb}%
  \BibitemOpen
  \bibfield  {author} {\bibinfo {author} {\bibfnamefont {R.~E.}\ \bibnamefont
  {Peierls}},\ }\href {https://doi.org/10.1098/rspa.1952.0158} {\bibfield
  {journal} {\bibinfo  {journal} {Proc. Roy. Soc. Lond. A}\ }\textbf {\bibinfo
  {volume} {214}},\ \bibinfo {pages} {143} (\bibinfo {year}
  {1952})}\BibitemShut {NoStop}%
\bibitem [{\citenamefont {Adamo}\ \emph {et~al.}(2022)\citenamefont {Adamo},
  \citenamefont {Cristofoli},\ and\ \citenamefont {Tourkine}}]{Adamo:2021rfq}%
  \BibitemOpen
  \bibfield  {author} {\bibinfo {author} {\bibfnamefont {T.}~\bibnamefont
  {Adamo}}, \bibinfo {author} {\bibfnamefont {A.}~\bibnamefont {Cristofoli}},\
  and\ \bibinfo {author} {\bibfnamefont {P.}~\bibnamefont {Tourkine}},\ }\href
  {https://doi.org/10.21468/SciPostPhys.13.2.032} {\bibfield  {journal}
  {\bibinfo  {journal} {SciPost Phys.}\ }\textbf {\bibinfo {volume} {13}},\
  \bibinfo {pages} {032} (\bibinfo {year} {2022})},\ \Eprint
  {https://arxiv.org/abs/2112.09113} {arXiv:2112.09113 [hep-th]} \BibitemShut
  {NoStop}%
\bibitem [{Note2()}]{Note2}%
  \BibitemOpen
  \bibinfo {note} {The positive frequency expansion $f(x) = \DOTSI \intop
  \ilimits@ _k e^{-ikx} f(k)$ and outgoing convention (reading coefficients of
  $f(-k)$ as vertex rules) are employed to ensure consistency with the Feynman
  rules given in Ref.~\cite {Mogull:2020sak}.}\BibitemShut {Stop}%
\bibitem [{Note3()}]{Note3}%
  \BibitemOpen
  \bibinfo {note} {A careful reader will argue that the LO contribution is the
  ``disconnected'' terms of order ${\protect \mathcal O}(G^{1/2})$
  corresponding to the Coulomb background generated by free particles. This
  contribution has been neglected since it does not seem to affect the
  dynamics, at least at the LO. See the footnote before Eq.~\protect \textup
  {\hbox {\mathsurround \z@ \protect \normalfont (\ignorespaces \ref
  {eq:LOradJ}\unskip \@@italiccorr )}} for possible violations of this
  assumption.}\BibitemShut {Stop}%
\bibitem [{Note4()}]{Note4}%
  \BibitemOpen
  \bibinfo {note} {The d'Alembertian operator $\square \Leftrightarrow - k^2$
  has an extra minus sign in momentum space.}\BibitemShut {Stop}%
\bibitem [{\citenamefont {Jakobsen}\ \emph {et~al.}(2021)\citenamefont
  {Jakobsen}, \citenamefont {Mogull}, \citenamefont {Plefka},\ and\
  \citenamefont {Steinhoff}}]{Jakobsen:2021smu}%
  \BibitemOpen
  \bibfield  {author} {\bibinfo {author} {\bibfnamefont {G.~U.}\ \bibnamefont
  {Jakobsen}}, \bibinfo {author} {\bibfnamefont {G.}~\bibnamefont {Mogull}},
  \bibinfo {author} {\bibfnamefont {J.}~\bibnamefont {Plefka}},\ and\ \bibinfo
  {author} {\bibfnamefont {J.}~\bibnamefont {Steinhoff}},\ }\href
  {https://doi.org/10.1103/PhysRevLett.126.201103} {\bibfield  {journal}
  {\bibinfo  {journal} {Phys. Rev. Lett.}\ }\textbf {\bibinfo {volume} {126}},\
  \bibinfo {pages} {201103} (\bibinfo {year} {2021})},\ \Eprint
  {https://arxiv.org/abs/2101.12688} {arXiv:2101.12688 [gr-qc]} \BibitemShut
  {NoStop}%
\bibitem [{Note5()}]{Note5}%
  \BibitemOpen
  \bibinfo {note} {There is a distinction made between the $i0^+$ prescription
  and the causality prescription. The $i0^+$ prescription can be retarded,
  advanced, or Feynman; the causality prescription can only be retarded or
  advanced.}\BibitemShut {Stop}%
\bibitem [{\citenamefont {Caron-Huot}\ \emph {et~al.}(2024)\citenamefont
  {Caron-Huot}, \citenamefont {Giroux}, \citenamefont {Hannesdottir},\ and\
  \citenamefont {Mizera}}]{Caron-Huot:2023vxl}%
  \BibitemOpen
  \bibfield  {author} {\bibinfo {author} {\bibfnamefont {S.}~\bibnamefont
  {Caron-Huot}}, \bibinfo {author} {\bibfnamefont {M.}~\bibnamefont {Giroux}},
  \bibinfo {author} {\bibfnamefont {H.~S.}\ \bibnamefont {Hannesdottir}},\ and\
  \bibinfo {author} {\bibfnamefont {S.}~\bibnamefont {Mizera}},\ }\href
  {https://doi.org/10.1007/JHEP01(2024)139} {\bibfield  {journal} {\bibinfo
  {journal} {JHEP}\ }\textbf {\bibinfo {volume} {01}},\ \bibinfo {pages}
  {139}},\ \Eprint {https://arxiv.org/abs/2308.02125} {arXiv:2308.02125
  [hep-th]} \BibitemShut {NoStop}%
\bibitem [{\citenamefont {Georgoudis}\ \emph
  {et~al.}(2024{\natexlab{b}})\citenamefont {Georgoudis}, \citenamefont
  {Heissenberg},\ and\ \citenamefont {Russo}}]{Georgoudis:2024pdz}%
  \BibitemOpen
  \bibfield  {author} {\bibinfo {author} {\bibfnamefont {A.}~\bibnamefont
  {Georgoudis}}, \bibinfo {author} {\bibfnamefont {C.}~\bibnamefont
  {Heissenberg}},\ and\ \bibinfo {author} {\bibfnamefont {R.}~\bibnamefont
  {Russo}},\ }\href {https://doi.org/10.1103/PhysRevD.109.106020} {\bibfield
  {journal} {\bibinfo  {journal} {Phys. Rev. D}\ }\textbf {\bibinfo {volume}
  {109}},\ \bibinfo {pages} {106020} (\bibinfo {year} {2024}{\natexlab{b}})},\
  \Eprint {https://arxiv.org/abs/2402.06361} {arXiv:2402.06361 [hep-th]}
  \BibitemShut {NoStop}%
\bibitem [{\citenamefont {Bini}\ \emph {et~al.}(2024)\citenamefont {Bini},
  \citenamefont {Damour}, \citenamefont {De~Angelis}, \citenamefont {Geralico},
  \citenamefont {Herderschee}, \citenamefont {Roiban},\ and\ \citenamefont
  {Teng}}]{Bini:2024rsy}%
  \BibitemOpen
  \bibfield  {author} {\bibinfo {author} {\bibfnamefont {D.}~\bibnamefont
  {Bini}}, \bibinfo {author} {\bibfnamefont {T.}~\bibnamefont {Damour}},
  \bibinfo {author} {\bibfnamefont {S.}~\bibnamefont {De~Angelis}}, \bibinfo
  {author} {\bibfnamefont {A.}~\bibnamefont {Geralico}}, \bibinfo {author}
  {\bibfnamefont {A.}~\bibnamefont {Herderschee}}, \bibinfo {author}
  {\bibfnamefont {R.}~\bibnamefont {Roiban}},\ and\ \bibinfo {author}
  {\bibfnamefont {F.}~\bibnamefont {Teng}},\ }\href
  {https://doi.org/10.1103/PhysRevD.109.125008} {\bibfield  {journal} {\bibinfo
   {journal} {Phys. Rev. D}\ }\textbf {\bibinfo {volume} {109}},\ \bibinfo
  {pages} {125008} (\bibinfo {year} {2024})},\ \Eprint
  {https://arxiv.org/abs/2402.06604} {arXiv:2402.06604 [hep-th]} \BibitemShut
  {NoStop}%
\bibitem [{\citenamefont {Cristofoli}\ \emph {et~al.}(2022)\citenamefont
  {Cristofoli}, \citenamefont {Gonzo}, \citenamefont {Kosower},\ and\
  \citenamefont {O'Connell}}]{Cristofoli:2021vyo}%
  \BibitemOpen
  \bibfield  {author} {\bibinfo {author} {\bibfnamefont {A.}~\bibnamefont
  {Cristofoli}}, \bibinfo {author} {\bibfnamefont {R.}~\bibnamefont {Gonzo}},
  \bibinfo {author} {\bibfnamefont {D.~A.}\ \bibnamefont {Kosower}},\ and\
  \bibinfo {author} {\bibfnamefont {D.}~\bibnamefont {O'Connell}},\ }\href
  {https://doi.org/10.1103/PhysRevD.106.056007} {\bibfield  {journal} {\bibinfo
   {journal} {Phys. Rev. D}\ }\textbf {\bibinfo {volume} {106}},\ \bibinfo
  {pages} {056007} (\bibinfo {year} {2022})},\ \Eprint
  {https://arxiv.org/abs/2107.10193} {arXiv:2107.10193 [hep-th]} \BibitemShut
  {NoStop}%
\bibitem [{\citenamefont {Brandhuber}\ \emph {et~al.}(2023)\citenamefont
  {Brandhuber}, \citenamefont {Brown}, \citenamefont {Chen}, \citenamefont
  {De~Angelis}, \citenamefont {Gowdy},\ and\ \citenamefont
  {Travaglini}}]{Brandhuber:2023hhy}%
  \BibitemOpen
  \bibfield  {author} {\bibinfo {author} {\bibfnamefont {A.}~\bibnamefont
  {Brandhuber}}, \bibinfo {author} {\bibfnamefont {G.~R.}\ \bibnamefont
  {Brown}}, \bibinfo {author} {\bibfnamefont {G.}~\bibnamefont {Chen}},
  \bibinfo {author} {\bibfnamefont {S.}~\bibnamefont {De~Angelis}}, \bibinfo
  {author} {\bibfnamefont {J.}~\bibnamefont {Gowdy}},\ and\ \bibinfo {author}
  {\bibfnamefont {G.}~\bibnamefont {Travaglini}},\ }\href
  {https://doi.org/10.1007/JHEP06(2023)048} {\bibfield  {journal} {\bibinfo
  {journal} {JHEP}\ }\textbf {\bibinfo {volume} {06}},\ \bibinfo {pages}
  {048}},\ \Eprint {https://arxiv.org/abs/2303.06111} {arXiv:2303.06111
  [hep-th]} \BibitemShut {NoStop}%
\bibitem [{\citenamefont {Herderschee}\ \emph {et~al.}(2023)\citenamefont
  {Herderschee}, \citenamefont {Roiban},\ and\ \citenamefont
  {Teng}}]{Herderschee:2023fxh}%
  \BibitemOpen
  \bibfield  {author} {\bibinfo {author} {\bibfnamefont {A.}~\bibnamefont
  {Herderschee}}, \bibinfo {author} {\bibfnamefont {R.}~\bibnamefont
  {Roiban}},\ and\ \bibinfo {author} {\bibfnamefont {F.}~\bibnamefont {Teng}},\
  }\href {https://doi.org/10.1007/JHEP06(2023)004} {\bibfield  {journal}
  {\bibinfo  {journal} {JHEP}\ }\textbf {\bibinfo {volume} {06}},\ \bibinfo
  {pages} {004}},\ \Eprint {https://arxiv.org/abs/2303.06112} {arXiv:2303.06112
  [hep-th]} \BibitemShut {NoStop}%
\bibitem [{\citenamefont {Elkhidir}\ \emph {et~al.}(2024)\citenamefont
  {Elkhidir}, \citenamefont {O'Connell}, \citenamefont {Sergola},\ and\
  \citenamefont {Vazquez-Holm}}]{Elkhidir:2023dco}%
  \BibitemOpen
  \bibfield  {author} {\bibinfo {author} {\bibfnamefont {A.}~\bibnamefont
  {Elkhidir}}, \bibinfo {author} {\bibfnamefont {D.}~\bibnamefont {O'Connell}},
  \bibinfo {author} {\bibfnamefont {M.}~\bibnamefont {Sergola}},\ and\ \bibinfo
  {author} {\bibfnamefont {I.~A.}\ \bibnamefont {Vazquez-Holm}},\ }\href
  {https://doi.org/10.1007/JHEP07(2024)272} {\bibfield  {journal} {\bibinfo
  {journal} {JHEP}\ }\textbf {\bibinfo {volume} {07}},\ \bibinfo {pages}
  {272}},\ \Eprint {https://arxiv.org/abs/2303.06211} {arXiv:2303.06211
  [hep-th]} \BibitemShut {NoStop}%
\bibitem [{\citenamefont {Georgoudis}\ \emph {et~al.}(2023)\citenamefont
  {Georgoudis}, \citenamefont {Heissenberg},\ and\ \citenamefont
  {Vazquez-Holm}}]{Georgoudis:2023lgf}%
  \BibitemOpen
  \bibfield  {author} {\bibinfo {author} {\bibfnamefont {A.}~\bibnamefont
  {Georgoudis}}, \bibinfo {author} {\bibfnamefont {C.}~\bibnamefont
  {Heissenberg}},\ and\ \bibinfo {author} {\bibfnamefont {I.}~\bibnamefont
  {Vazquez-Holm}},\ }\href {https://doi.org/10.1007/JHEP06(2023)126} {\bibfield
   {journal} {\bibinfo  {journal} {JHEP}\ }\textbf {\bibinfo {volume}
  {2023}}\bibfield  {number} {\bibinfo  {number} { (06)},\ \bibinfo {pages}
  {126}},\ }\Eprint {https://arxiv.org/abs/2303.07006} {arXiv:2303.07006
  [hep-th]} \BibitemShut {NoStop}%
\bibitem [{\citenamefont {Bini}\ \emph {et~al.}(2023)\citenamefont {Bini},
  \citenamefont {Damour},\ and\ \citenamefont {Geralico}}]{Bini:2023fiz}%
  \BibitemOpen
  \bibfield  {author} {\bibinfo {author} {\bibfnamefont {D.}~\bibnamefont
  {Bini}}, \bibinfo {author} {\bibfnamefont {T.}~\bibnamefont {Damour}},\ and\
  \bibinfo {author} {\bibfnamefont {A.}~\bibnamefont {Geralico}},\ }\href
  {https://doi.org/10.1103/PhysRevD.108.124052} {\bibfield  {journal} {\bibinfo
   {journal} {Phys. Rev. D}\ }\textbf {\bibinfo {volume} {108}},\ \bibinfo
  {pages} {124052} (\bibinfo {year} {2023})},\ \Eprint
  {https://arxiv.org/abs/2309.14925} {arXiv:2309.14925 [gr-qc]} \BibitemShut
  {NoStop}%
\bibitem [{Note6()}]{Note6}%
  \BibitemOpen
  \bibinfo {note} {For spinning systems, $\DOTSB \sum@ \slimits@ _i \{ \chi
  _{(1)} , S_i^{\m \n } \} \protect \frac {\partial }{\partial S_i^{\m \n }}$
  implements precession by the LO spin kick, consistent with the conventional
  picture of saddle-point shifts~\cite {Luna:2023uwd}. In other words, the
  $\protect \mathcal {D}_{SL}$ operator introduced in Ref.~\cite {Bern:2020buy}
  can be viewed as compensation terms for neglecting nested brackets in the
  scattering generator equation [Eq.\protect \textup {\hbox {\mathsurround \z@
  \protect \normalfont (\ignorespaces \ref {eq:scgen_eq}\unskip \@@italiccorr
  )}}] and ``noncommutativity'' of the impact parameter space [Eq.\protect
  \textup {\hbox {\mathsurround \z@ \protect \normalfont (\ignorespaces \ref
  {eq:bbPB}\unskip \@@italiccorr )}}]; see e.g. the scattering observable
  computations of Ref.~\cite {Akpinar:2025bkt} where instead of defining $b^\m
  $ to manifestly satisfy the orthogonality conditions $b \cdot P_i = 0$
  [Eq.\protect \textup {\hbox {\mathsurround \z@ \protect \normalfont
  (\ignorespaces \ref {eq:impbdef}\unskip \@@italiccorr )}}], the Dirac
  brackets constructed from imposing orthogonality conditions as constraints
  were used. When brackets between variables that manifestly satisfy the
  constraint conditions are considered, Poisson brackets can be used in place
  of Dirac brackets~\cite {Kim:2024grz}.}\BibitemShut {Stop}%
\bibitem [{\citenamefont {Herrmann}\ \emph {et~al.}(2021)\citenamefont
  {Herrmann}, \citenamefont {Parra-Martinez}, \citenamefont {Ruf},\ and\
  \citenamefont {Zeng}}]{Herrmann:2021lqe}%
  \BibitemOpen
  \bibfield  {author} {\bibinfo {author} {\bibfnamefont {E.}~\bibnamefont
  {Herrmann}}, \bibinfo {author} {\bibfnamefont {J.}~\bibnamefont
  {Parra-Martinez}}, \bibinfo {author} {\bibfnamefont {M.~S.}\ \bibnamefont
  {Ruf}},\ and\ \bibinfo {author} {\bibfnamefont {M.}~\bibnamefont {Zeng}},\
  }\href {https://doi.org/10.1103/PhysRevLett.126.201602} {\bibfield  {journal}
  {\bibinfo  {journal} {Phys. Rev. Lett.}\ }\textbf {\bibinfo {volume} {126}},\
  \bibinfo {pages} {201602} (\bibinfo {year} {2021})},\ \Eprint
  {https://arxiv.org/abs/2101.07255} {arXiv:2101.07255 [hep-th]} \BibitemShut
  {NoStop}%
\bibitem [{Note7()}]{Note7}%
  \BibitemOpen
  \bibinfo {note} {This relation follows from translation invariance of LO
  radiation eikonal $\chi _{(1.5)}^{H^1} [H(x)]$ in position space, where $x^\m
  $ is dual to $k^\m $ in Eq.\protect \textup {\hbox {\mathsurround \z@
  \protect \normalfont (\ignorespaces \ref {eq:LOradeik}\unskip \@@italiccorr
  )}}; $\protect \frac {\partial \chi _{(1.5)}^{H^1}}{\partial X_1^\m } +
  \protect \frac {\partial \chi _{(1.5)}^{H^1}}{\partial X_2^\m } + \protect
  \frac {\partial \chi _{(1.5)}^{H^1}}{\partial x^\m } = 0$. When reformulated
  in position space, the integrand of Eq.\protect \textup {\hbox {\mathsurround
  \z@ \protect \normalfont (\ignorespaces \ref {eq:LOradmom}\unskip
  \@@italiccorr )}} is equivalent to the integrand of Eq.\protect \textup
  {\hbox {\mathsurround \z@ \protect \normalfont (\ignorespaces \ref
  {eq:3PMrl2}\unskip \@@italiccorr )}} with $\protect \frac {\partial
  }{\partial X_{1\m }}$ substituted by $\protect \frac {\partial }{\partial
  x_\m }$, and translation invariance implies $P_{H, \protect \text {out}}^\m +
  \D _{(3,rl)} P_1^\m + \D _{(3,rl)} P_2^\m = 0$.}\BibitemShut {Stop}%
\bibitem [{\citenamefont {Jakobsen}(2023)}]{Jakobsen:2023oow}%
  \BibitemOpen
  \bibfield  {author} {\bibinfo {author} {\bibfnamefont {G.~U.}\ \bibnamefont
  {Jakobsen}},\ }\emph {\bibinfo {title} {{Gravitational Scattering of Compact
  Bodies from Worldline Quantum Field Theory}}},\ \href
  {https://doi.org/10.18452/27075} {Ph.D. thesis},\ \bibinfo  {school}
  {Humboldt U., Berlin, Humboldt U., Berlin (main)} (\bibinfo {year} {2023}),\
  \Eprint {https://arxiv.org/abs/2308.04388} {arXiv:2308.04388 [hep-th]}
  \BibitemShut {NoStop}%
\bibitem [{\citenamefont {Heissenberg}(2025)}]{Heissenberg:2025ocy}%
  \BibitemOpen
  \bibfield  {author} {\bibinfo {author} {\bibfnamefont {C.}~\bibnamefont
  {Heissenberg}},\ }\href@noop {} {\  (\bibinfo {year} {2025})},\ \Eprint
  {https://arxiv.org/abs/2501.02904} {arXiv:2501.02904 [hep-th]} \BibitemShut
  {NoStop}%
\bibitem [{Note8()}]{Note8}%
  \BibitemOpen
  \bibinfo {note} {Eq.~\protect \textup {\hbox {\mathsurround \z@ \protect
  \normalfont (\ignorespaces \ref {eq:LOradJ}\unskip \@@italiccorr )}} assumes
  vanishing contributions from the ``disconnected'' radiation eikonal $\chi
  _{(0.5)}^{H^1}$. The LO will become ${\protect \mathcal O}(G^2)$ instead if
  this assumption is violated.}\BibitemShut {Stop}%
\bibitem [{\citenamefont {Manohar}\ \emph {et~al.}(2022)\citenamefont
  {Manohar}, \citenamefont {Ridgway},\ and\ \citenamefont
  {Shen}}]{Manohar:2022dea}%
  \BibitemOpen
  \bibfield  {author} {\bibinfo {author} {\bibfnamefont {A.~V.}\ \bibnamefont
  {Manohar}}, \bibinfo {author} {\bibfnamefont {A.~K.}\ \bibnamefont
  {Ridgway}},\ and\ \bibinfo {author} {\bibfnamefont {C.-H.}\ \bibnamefont
  {Shen}},\ }\href {https://doi.org/10.1103/PhysRevLett.129.121601} {\bibfield
  {journal} {\bibinfo  {journal} {Phys. Rev. Lett.}\ }\textbf {\bibinfo
  {volume} {129}},\ \bibinfo {pages} {121601} (\bibinfo {year} {2022})},\
  \Eprint {https://arxiv.org/abs/2203.04283} {arXiv:2203.04283 [hep-th]}
  \BibitemShut {NoStop}%
\bibitem [{\citenamefont {Di~Vecchia}\ \emph
  {et~al.}(2022{\natexlab{b}})\citenamefont {Di~Vecchia}, \citenamefont
  {Heissenberg},\ and\ \citenamefont {Russo}}]{DiVecchia:2022owy}%
  \BibitemOpen
  \bibfield  {author} {\bibinfo {author} {\bibfnamefont {P.}~\bibnamefont
  {Di~Vecchia}}, \bibinfo {author} {\bibfnamefont {C.}~\bibnamefont
  {Heissenberg}},\ and\ \bibinfo {author} {\bibfnamefont {R.}~\bibnamefont
  {Russo}},\ }\href {https://doi.org/10.1007/JHEP08(2022)172} {\bibfield
  {journal} {\bibinfo  {journal} {JHEP}\ }\textbf {\bibinfo {volume} {08}},\
  \bibinfo {pages} {172}},\ \Eprint {https://arxiv.org/abs/2203.11915}
  {arXiv:2203.11915 [hep-th]} \BibitemShut {NoStop}%
\bibitem [{\citenamefont {Mao}\ and\ \citenamefont {Zeng}(2025)}]{Mao:2024ryq}%
  \BibitemOpen
  \bibfield  {author} {\bibinfo {author} {\bibfnamefont {P.}~\bibnamefont
  {Mao}}\ and\ \bibinfo {author} {\bibfnamefont {B.}~\bibnamefont {Zeng}},\
  }\href {https://doi.org/10.1103/PhysRevD.111.L021502} {\bibfield  {journal}
  {\bibinfo  {journal} {Phys. Rev. D}\ }\textbf {\bibinfo {volume} {111}},\
  \bibinfo {pages} {L021502} (\bibinfo {year} {2025})},\ \Eprint
  {https://arxiv.org/abs/2406.07943} {arXiv:2406.07943 [gr-qc]} \BibitemShut
  {NoStop}%
\bibitem [{\citenamefont {Biswas}\ and\ \citenamefont
  {Parra-Martinez}(2024)}]{Biswas:2024ept}%
  \BibitemOpen
  \bibfield  {author} {\bibinfo {author} {\bibfnamefont {S.}~\bibnamefont
  {Biswas}}\ and\ \bibinfo {author} {\bibfnamefont {J.}~\bibnamefont
  {Parra-Martinez}},\ }\href@noop {} {\  (\bibinfo {year} {2024})},\ \Eprint
  {https://arxiv.org/abs/2411.09016} {arXiv:2411.09016 [hep-th]} \BibitemShut
  {NoStop}%
\bibitem [{\citenamefont {Cristofoli}\ \emph {et~al.}(2024)\citenamefont
  {Cristofoli}, \citenamefont {Gonzo}, \citenamefont {Moynihan}, \citenamefont
  {O'Connell}, \citenamefont {Ross}, \citenamefont {Sergola},\ and\
  \citenamefont {White}}]{Cristofoli:2021jas}%
  \BibitemOpen
  \bibfield  {author} {\bibinfo {author} {\bibfnamefont {A.}~\bibnamefont
  {Cristofoli}}, \bibinfo {author} {\bibfnamefont {R.}~\bibnamefont {Gonzo}},
  \bibinfo {author} {\bibfnamefont {N.}~\bibnamefont {Moynihan}}, \bibinfo
  {author} {\bibfnamefont {D.}~\bibnamefont {O'Connell}}, \bibinfo {author}
  {\bibfnamefont {A.}~\bibnamefont {Ross}}, \bibinfo {author} {\bibfnamefont
  {M.}~\bibnamefont {Sergola}},\ and\ \bibinfo {author} {\bibfnamefont {C.~D.}\
  \bibnamefont {White}},\ }\href {https://doi.org/10.1007/JHEP06(2024)181}
  {\bibfield  {journal} {\bibinfo  {journal} {JHEP}\ }\textbf {\bibinfo
  {volume} {06}},\ \bibinfo {pages} {181}},\ \Eprint
  {https://arxiv.org/abs/2112.07556} {arXiv:2112.07556 [hep-th]} \BibitemShut
  {NoStop}%
\bibitem [{\citenamefont {Shapiro}(1964)}]{Shapiro:1964uw}%
  \BibitemOpen
  \bibfield  {author} {\bibinfo {author} {\bibfnamefont {I.~I.}\ \bibnamefont
  {Shapiro}},\ }\href {https://doi.org/10.1103/PhysRevLett.13.789} {\bibfield
  {journal} {\bibinfo  {journal} {Phys. Rev. Lett.}\ }\textbf {\bibinfo
  {volume} {13}},\ \bibinfo {pages} {789} (\bibinfo {year} {1964})}\BibitemShut
  {NoStop}%
\bibitem [{\citenamefont {Camanho}\ \emph {et~al.}(2016)\citenamefont
  {Camanho}, \citenamefont {Edelstein}, \citenamefont {Maldacena},\ and\
  \citenamefont {Zhiboedov}}]{Camanho:2014apa}%
  \BibitemOpen
  \bibfield  {author} {\bibinfo {author} {\bibfnamefont {X.~O.}\ \bibnamefont
  {Camanho}}, \bibinfo {author} {\bibfnamefont {J.~D.}\ \bibnamefont
  {Edelstein}}, \bibinfo {author} {\bibfnamefont {J.}~\bibnamefont
  {Maldacena}},\ and\ \bibinfo {author} {\bibfnamefont {A.}~\bibnamefont
  {Zhiboedov}},\ }\href {https://doi.org/10.1007/JHEP02(2016)020} {\bibfield
  {journal} {\bibinfo  {journal} {JHEP}\ }\textbf {\bibinfo {volume} {02}},\
  \bibinfo {pages} {020}},\ \Eprint {https://arxiv.org/abs/1407.5597}
  {arXiv:1407.5597 [hep-th]} \BibitemShut {NoStop}%
\bibitem [{\citenamefont {Chen}\ \emph {et~al.}(2022)\citenamefont {Chen},
  \citenamefont {Chung}, \citenamefont {Huang},\ and\ \citenamefont
  {Kim}}]{Chen:2022clh}%
  \BibitemOpen
  \bibfield  {author} {\bibinfo {author} {\bibfnamefont {W.-M.}\ \bibnamefont
  {Chen}}, \bibinfo {author} {\bibfnamefont {M.-Z.}\ \bibnamefont {Chung}},
  \bibinfo {author} {\bibfnamefont {Y.-t.}\ \bibnamefont {Huang}},\ and\
  \bibinfo {author} {\bibfnamefont {J.-W.}\ \bibnamefont {Kim}},\ }\href
  {https://doi.org/10.1007/JHEP12(2022)058} {\bibfield  {journal} {\bibinfo
  {journal} {JHEP}\ }\textbf {\bibinfo {volume} {12}},\ \bibinfo {pages}
  {058}},\ \Eprint {https://arxiv.org/abs/2205.07305} {arXiv:2205.07305
  [hep-th]} \BibitemShut {NoStop}%
\bibitem [{\citenamefont {Kim}(2022)}]{Kim:2022iub}%
  \BibitemOpen
  \bibfield  {author} {\bibinfo {author} {\bibfnamefont {J.-W.}\ \bibnamefont
  {Kim}},\ }\href {https://doi.org/10.1103/PhysRevD.106.L081901} {\bibfield
  {journal} {\bibinfo  {journal} {Phys. Rev. D}\ }\textbf {\bibinfo {volume}
  {106}},\ \bibinfo {pages} {L081901} (\bibinfo {year} {2022})},\ \Eprint
  {https://arxiv.org/abs/2207.04970} {arXiv:2207.04970 [hep-th]} \BibitemShut
  {NoStop}%
\bibitem [{\citenamefont {Bern}\ \emph
  {et~al.}(2021{\natexlab{b}})\citenamefont {Bern}, \citenamefont {Luna},
  \citenamefont {Roiban}, \citenamefont {Shen},\ and\ \citenamefont
  {Zeng}}]{Bern:2020buy}%
  \BibitemOpen
  \bibfield  {author} {\bibinfo {author} {\bibfnamefont {Z.}~\bibnamefont
  {Bern}}, \bibinfo {author} {\bibfnamefont {A.}~\bibnamefont {Luna}}, \bibinfo
  {author} {\bibfnamefont {R.}~\bibnamefont {Roiban}}, \bibinfo {author}
  {\bibfnamefont {C.-H.}\ \bibnamefont {Shen}},\ and\ \bibinfo {author}
  {\bibfnamefont {M.}~\bibnamefont {Zeng}},\ }\href
  {https://doi.org/10.1103/PhysRevD.104.065014} {\bibfield  {journal} {\bibinfo
   {journal} {Phys. Rev. D}\ }\textbf {\bibinfo {volume} {104}},\ \bibinfo
  {pages} {065014} (\bibinfo {year} {2021}{\natexlab{b}})},\ \Eprint
  {https://arxiv.org/abs/2005.03071} {arXiv:2005.03071 [hep-th]} \BibitemShut
  {NoStop}%
\bibitem [{\citenamefont {Akpinar}\ \emph {et~al.}(2025)\citenamefont
  {Akpinar}, \citenamefont {Febres~Cordero}, \citenamefont {Kraus},
  \citenamefont {Smirnov},\ and\ \citenamefont {Zeng}}]{Akpinar:2025bkt}%
  \BibitemOpen
  \bibfield  {author} {\bibinfo {author} {\bibfnamefont {D.}~\bibnamefont
  {Akpinar}}, \bibinfo {author} {\bibfnamefont {F.}~\bibnamefont
  {Febres~Cordero}}, \bibinfo {author} {\bibfnamefont {M.}~\bibnamefont
  {Kraus}}, \bibinfo {author} {\bibfnamefont {A.}~\bibnamefont {Smirnov}},\
  and\ \bibinfo {author} {\bibfnamefont {M.}~\bibnamefont {Zeng}},\ }\href@noop
  {} {\  (\bibinfo {year} {2025})},\ \Eprint {https://arxiv.org/abs/2502.08961}
  {arXiv:2502.08961 [hep-th]} \BibitemShut {NoStop}%
\end{thebibliography}%

%

\end{document}